\documentclass[english,12pt, draftclsnofoot, onecolumn]{IEEEtran}
\usepackage[T1]{fontenc}
\usepackage[latin9]{inputenc}
\usepackage{color}
\usepackage{babel}
\usepackage{float}
\usepackage{amsmath}
\usepackage{amsthm}
\usepackage{amssymb}
\usepackage{stackrel}
\usepackage{graphicx}
\usepackage{esint}
\usepackage[unicode=true,
 bookmarks=true,bookmarksnumbered=false,bookmarksopen=false,
 breaklinks=false,pdfborder={0 0 1},backref=false,colorlinks=true]
 {hyperref}

\makeatletter

\providecommand{\tabularnewline}{\\}
\floatstyle{ruled}
\newfloat{algorithm}{tbp}{loa}
\providecommand{\algorithmname}{Algorithm}
\floatname{algorithm}{\protect\algorithmname}

\theoremstyle{plain}
\newtheorem{lem}{\protect\lemmaname}
\theoremstyle{plain}
\newtheorem{thm}{\protect\theoremname}
\theoremstyle{remark}
\newtheorem{rem}{\protect\remarkname}
\theoremstyle{plain}
\newtheorem{cor}{\protect\corollaryname}

\usepackage{babel}
\usepackage{cite}
\usepackage{algorithm}
\usepackage{algorithmic}
\usepackage[colorlinks=true]{hyperref}
\usepackage{graphicx}
\usepackage{esint}

\@ifundefined{showcaptionsetup}{}{%
 \PassOptionsToPackage{caption=false}{subfig}}
\usepackage{subfig}
\makeatother

\providecommand{\corollaryname}{Corollary}
\providecommand{\lemmaname}{Lemma}
\providecommand{\remarkname}{Remark}
\providecommand{\theoremname}{Theorem}

\begin{document}
\title{Fluid Antenna System: New Insights on Outage Probability and Diversity
Gain\thanks{The work of W. K. New, K.-K. Wong and K.-F. Tong is supported by the Engineering and Physical Sciences Research Council (EPSRC) under grant EP/W026813/1. The work of C.-B. Chae is supported by the Institute for Information and Communication Technology Promotion (IITP) grant funded by the Ministry of Science and ICT (MSIT), Korea (No. 2021-0-02208, No. 2021-0-00486). The work of H. Xu is supported by the European Union's Horizon 2020 Research and Innovation Programme under Marie Sklodowska-Curie Grant No. 101024636.}}
\author{Wee Kiat New, \textit{Member, IEEE}, Kai-Kit Wong, \textit{Fellow,
IEEE}, \\
Hao Xu, \textit{Member, IEEE,} Kin-Fai Tong, \textit{Fellow, IEEE}, \\
\textit{and} Chan-Byoung Chae, \textit{Fellow, IEEE}\thanks{Wee Kiat New (email: a.new@ucl.ac.uk), Kai-Kit Wong (email: kai-kit.wong@ucl.ac.uk),
Hao Xu (email:hao.xu@ucl.uk), and Kin-Fai Tong (email: k.tong@ucl.ac.uk)
are with the Department of Electronic and Electrical Engineering,
University College London, London WC1E 6BT, United Kingdom.}\thanks{Chan-Byoung Chae (email: cbchae@yonsei.ac.kr) is with the School of Integrated Technology, Yonsei University, Seoul 03722 Korea. Kai-Kit Wong is also affiliated with Yonsei Frontier Lab., Yonsei University, Seoul 03722, Korea.}}

\maketitle
\vspace{-1cm}

\begin{abstract}
To enable innovative applications and services, both industry and
academia are exploring new technologies for sixth generation (6G)
communications. One of the promising candidates is fluid antenna system
(FAS). Unlike existing systems, FAS is a novel communication technology
where its antenna can freely change its position and shape within
a given space. Compared to the traditional systems, this unique capability
has the potential of providing higher diversity and interference-free
communications. Nevertheless, the performance limits of FAS remain
unclear as its system properties are difficult to analyze. To address
this, we approximate the outage probability and diversity gain of
FAS in closed-form expressions. We then propose a suboptimal FAS with
$N^{*}$ ports, where a significant gain can be obtained over FAS
with $N^{*}-1$ ports whilst FAS with $N^{*}+1$ ports only yields
marginal improvement over the proposed suboptimal FAS. In this paper,
we also provide analytical and simulation results to unfold the key
factors that affect the performance of FAS. Limited to systems with
one active radio frequency (RF)-chain, we show that the proposed suboptimal
FAS outperforms single-antenna (SISO) system and selection combining
(SC) system in terms of outage probability. Interestingly, when the
given space is $\frac{\lambda}{2}$, the outage probability of the
proposed suboptimal FAS with one active RF-chain achieves near to
that of the maximal ratio combining (MRC) system with multiple active
RF-chains. 
\end{abstract}

\begin{IEEEkeywords}
6G, fluid antenna system, outage probability, diversity gain, performance
analysis. 
\end{IEEEkeywords}

\section{Introduction}

Fifth generation (5G) wireless networks have recently been deployed
worldwide and thus the industry and academia are now looking for new
technologies to maximize the potentials of sixth generation (6G) wireless
networks. One of the promising candidates is fluid antenna system
(FAS). Unlike traditional antenna systems, FAS is a software-controllable
fluidic, conductive or dielectric structure that can freely adjust
its position and shape within a given space \cite{9770295}. The
most basic single fluid antenna consists of one radio frequency (RF)-chain
and $N$ preset positions (known as ports) that are distributed within
a given space while more advanced designs are also possible. The fluid
radiator can freely switch its position among these ports to obtain
a stronger channel gain, lower interference, and other desirable performance
\cite{wong2022bruce}.

Fluid antenna is now feasible thanks to the recent
advancement of using liquid metals and ionized solutions for antennas.
Some prototypes can be found in \cite{9899974,9899924,9539785,9331647}.
As discussed in \cite{9264694}, other flexible antenna structures
such as software-controlled pixel antenna or movable antenna can also
be considered as fluid antenna. In essence, the
key principle of FAS is to exploit its dynamic position and shape
to achieve ultimate diversity and multiplexing gains \cite{9264694}.
Moreover, in the future, FAS can be applied together with other 6G
candidates such as re-configurable intelligent surfaces (RIS), massive
multiple-input multiple-output (MIMO) and terahertz (THz) communications.
In particular, FAS can help to reduce the optimization complexity
of RIS \cite{9770295}, improve the multiplexing gain of massive MIMO
\cite{wong2022extra} and combat the high path loss effect of THz
communications \cite{tlebaldiyeva2023outage}.

Despite its advantages, the fundamental limits of FAS and key factors
that affect its performance remain unclear. One of the reasons is
because the channels of FAS are strongly correlated since the ports
can be closely placed to each other. Consequently, the probability
density function (PDF) and cumulative distribution function (CDF)
of FAS channels are intractable \cite{10103838}. As a result,
the outage probability and diversity gain of FAS are not known in
closed-form expressions. In addition, increasing the number of ports
of FAS has an inherit diminishing gain due to one active RF-chain
\cite{4065786}.\footnote{Throughout this paper, we refer to an active RF-chain as the RF-chain
used for communications. In contrast, the term RF-chains refers to
a collection of RF-chains that are connected to each antenna for it
to work as intended.} Thus, a suboptimal number of ports that are required to achieve a
satisfactory performance is not known. Yet, this number is practically
and theoretically important as it reduces the implementation challenges
and analysis complexity.

Conceptually, FAS can be viewed similar to a traditional
selection combining (SC) system since both systems use only one active
RF-chain and there is a set of ports (i.e., FAS) or antennas (i.e.,
SC) to select from. Nevertheless, unlike traditional SC system, FAS
can have infinitely many ports in a limited space (e.g., when using
liquid metals) which makes the implementation and analysis much more
challenging. In addition, the unique capability of freely switching
the radiating element among the ports can be exploited to mitigate
multi-user interference. These features are impractical in traditional
SC systems.

State-of-the-arts show that FAS outperforms maximal ratio combining
(MRC) system if the number of ports is sufficiently large \cite{9264694}.
In fact, \cite{9264694} proves that FAS achieves arbitrarily small
outage for a fixed rate/signal-to-noise ratio (SNR) as $N\rightarrow\infty$.
In \cite{9131873}, the authors reveal that the ergodic capacity of
FAS increases with $N$ and thus FAS can outperform MRC in terms of
ergodic capacity. Interestingly, FAS can also be used for multiple
access. Specifically, \cite{9650760} proposes a fluid antenna multiple
access (FAMA) system which leverages the moment of deep fades in space
to reduce multi-user interference. Motivated by these works, \cite{9838484}
employs stochastic geometry to analyze the outage probability of FAS
in large-scale downlink cellular networks and \cite{9771633} analyzes
the performance of FAS in a more general correlated fading channel.

Nevertheless, \cite{wong2022closed} alludes that the channel modeling
in the previous works might be inaccurate. To address this, \cite{10103838}
proposes a highly complicated channel model to follow closely the
spatial correlation of the Jake's model. Using this channel model,
they highlight that FAS has limited performance gain as $N$ increases.
Yet, the key reasons that limit the performance of FAS remain ambiguous.
This is because the eigenvalue and eigenvector entries that are used
in the analytical PDF/CDF expressions provide limited insights.

It is important to highlight that deriving the PDF/CDF of FAS channels
is extremely challenging \cite{10103838}. This is because
the channels of FAS are strongly correlated and thus they have to
be formulated in terms of multivariate correlated Rayleigh distributions.
Over the past few decades, extensive efforts have been dedicated to
this problem \cite{LE20191}. However, most of the works only obtain
the bivariate \cite{miller1969complex,634675}, trivariate \cite{miller1969complex,4711160,1556835},
or quadvariate \cite{1556835,tekinay2020moments} distributions while
other works restrict the correlation matrix to certain forms (e.g.,
equally correlated \cite{1356206} and exponentially correlated \cite{1221771}).
Fortunately, the multivariate PDF/CDF of arbitrarily correlated Rayleigh
distributions are recently derived in \cite{https://doi.org/10.1002/dac.3510,wiegand2019series,WIEGAND2020107664}.
Nevertheless, the assumption of non-singular correlation matrix is
retained. In this paper, we omit this assumption (i.e., our correlation
matrix could be near-singular) and address the computation problem
via a suboptimal approximation.\footnote{The computational problem of a near-singular correlation matrix is
much harder to address than that of a singular matrix. This is because
we can obtain an independent matrix from a singular matrix by removing
the dependent entries \cite{gallager2008principles}. But in the near-singular
case, this approach cannot be applied. Instead, we need to rely on
approximations.}

In addition to the above works, \cite{9715064} develops a port selection
algorithm that can approach the performance of optimal FAS when only
the received SNR of a few ports are observed. Furthermore, \cite{zhu2022modeling}
considers a field-response channel model while omitting the spatial
correlation effect and \cite{ma2022mimo} extends the model to a MIMO
scenario. Moreover, FAMA can be categorized into i) slow-FAMA and
ii) fast-FAMA. The earlier switches its port when the channel changes
\cite{Noor1} while the latter switches its port on a symbol-by-symbol
basis \cite{FastFAMA}. The analytical outage probability of two-user
FAMA is also derived in \cite{OutageFAMA}.

Motivated by the aforementioned works, this paper aims to understand
the fundamental limits of FAS as well as the key factors that affect
its performance. To this end, we approximate the outage probability
and diversity gain of FAS in closed-form expressions via a simple
and accurate channel model that follows closely the spatial correlation
of Jake's model. In addition, we propose a suboptimal FAS with $N^{*}$
ports as well as an algorithm to approximate $N^{*}$. The main contributions
of our paper are summarized as follows: 
\begin{itemize}
\item We employ a simple and accurate channel model that follows the spatial
correlation of Jake's model. Based on this channel model, we approximate
the outage probability in closed-form expressions. By applying Taylor
series approximation, we simplify the outage probability at high SNR
into a simpler and more meaningful expression. Using this result,
we obtain the diversity gain of FAS. 
\item We propose a suboptimal FAS with $N^{*}$ ports. The proposed suboptimal
FAS plays an important role as it enables FAS to achieve near-optimal
performance with minimal number of ports. In particular, one may define
$\varepsilon_{\text{tol}}$ to adjust the sub-optimality of the proposed
FAS. For example, if $\varepsilon_{\text{tol}}$ is small, the proposed
FAS is quantifiably near-optimal at a cost of more ports. In addition,
we develop a polynomial-time algorithm to approximate $N^{*}$. Besides,
$N^{*}$ can be used to address the near-singular correlation matrix
problem. 
\item We provide analytical and simulation results to demonstrate the key
parameters that affect the performance of FAS. Our discussions include
intuitive insights on the system characteristics as well as practical
guidelines for efficient FAS design. 
\end{itemize}
The rest of the paper is organized as follows: Section II details
the system model and performance metrics. Section III presents the
outage probability and diversity gain of FAS. The details of suboptimal
FAS and the algorithm to approximate $N^{*}$ are discussed in Section
IV. Section V provides our simulation results and we conclude the
paper in Section VI.

\textit{Notations:}\textsl{ }Scalar variables are denoted by italic letters (e.g., $c$), vectors are denoted by boldface italic small letters (e.g., $\boldsymbol{c}$) and matrices are denoted by boldface italic capital letters (e.g., $\boldsymbol{C}$). Besides, $\left(\cdot\right)^{T}$ denotes transpose, $\left(\cdot\right)^{H}$ denotes conjugate transpose, $\left(\cdot\right)^{-1}$ denotes inverse of a matrix while $\left|\cdot\right|$ and $\left\Vert \cdot\right\Vert _{F}$ denotes absolute and Frobenius norm, respectively. Throughout this paper, $\log(\cdot)$ denotes logarithm with base 2, $\mathbb{E}\left[\cdot\right]$ denotes the expectation and $\mathbb{P}\left\{ \cdot\right\} $ denotes the probability of an event. In addition, $f_{c}\left(\cdot\right)$ denotes the PDF of $c$, and $F_{c}\left(\cdot\right)$ denotes the CDF of $c$. The notation $\mathbf{\mathbb{\mathbf{1}}}_{c}\left\{ \cdot\right\}$ is an indication function for condition $c$ and $\left[\cdot\right]_{c}^{+/-}$ outputs the argument that is lower/upper bounded by $c$. To help readers with our mathematical content, the meanings of the key variables are explained in Table \ref{table1}.

\begin{table}
\centering{}\caption{\label{table1}The meanings of key variables}
\begin{tabular}{c||p{8cm}}
\hline 
Notation  & \centering {Meaning}\tabularnewline
\hline 
{$D_{\text{FAS}}$}  & {Diversity gain of FAS}\tabularnewline
\hline 
$h_{n}$  & Complex channel coefficient of the $n$-th port\tabularnewline
\hline 
$\hat{h}_{n}$  & Approximation of $h_{n}$\tabularnewline
\hline 
$\left|h_{\text{FAS}}\right|$  & Maximum signal envelope of FAS\tabularnewline
\hline 
$J_{m,n}$  & Spatial correlation between the $m$-th and $n$-th
ports\tabularnewline
\hline 
$\boldsymbol{J}$  & Spatial correlation matrix\tabularnewline
\hline 
$\boldsymbol{J}'$  & Spatial correlation matrix with $N\rightarrow\infty$\tabularnewline
\hline 
$\boldsymbol{K}$  & Co-factor of $\boldsymbol{J}$\tabularnewline
\hline 
$\lambda$  & Wavelength of the carrier frequency\tabularnewline
\hline 
$N$  & Total number of ports\tabularnewline
\hline 
$N'$  & Rank of $\boldsymbol{J}'$\tabularnewline
\hline 
$q$  & Minimum required rate\tabularnewline
\hline 
$SNR$  & Transmit SNR\tabularnewline
\hline 
$W$  & Length of the fluid antenna in terms of $\lambda$\tabularnewline
\hline 
$\Theta$  & Instantaneous received SNR of the receiver\tabularnewline
\hline 
\end{tabular}
\end{table}

\section{System Model}

In this paper, we consider a point-to-point FAS where the transmitter
is equipped with a conventional antenna and the receiver is equipped
with a fluid antenna. The fluid antenna consists
of one RF-chain and $N$ preset locations (also known as ports), which
are evenly distributed along a linear dimension of length $W\lambda$
where $\lambda$ is the wavelength of the carrier frequency. Since
the ports are closely packed together, there is a strong spatial correlation
among them. Based on Jake's model \cite{stuber1996principles}, the
spatial correlation between the $m$-th and $n$-th ports is given
by
\begin{equation}
J_{m,n}=\sigma^{2}J_{0}\left(2\pi\frac{\left(m-n\right)}{N-1}W\right),\label{eq:1}
\end{equation}
where $\sigma^{2}$ accounts for the large-scale fading effect and
$J_{0}\left(\cdot\right)$ is the zero-order Bessel function of the
first kind.

For ease of analysis, we introduce the correlation matrix $\boldsymbol{J}$
where 
\begin{equation}
\boldsymbol{J}=\left[\begin{array}{ccc}
J_{1,1} & \cdots & J_{1,N}\\
\vdots & \ddots & \vdots\\
J_{N,1} & \cdots & J_{N,N}
\end{array}\right].\label{eq:2}
\end{equation}
In (\ref{eq:2}), we have $J_{m,n}=J_{n,m}$. Therefore, using eigenvalue
decomposition, we can obtain $\boldsymbol{J}=\boldsymbol{U}\boldsymbol{\Lambda}\boldsymbol{U}^{H}$
where $\boldsymbol{U}$ is an $N\times N$ matrix whose $n$-th column
(denoted by $\boldsymbol{u}_{n}$) is the eigenvector of $\boldsymbol{J}$
and $\boldsymbol{\Lambda}=\text{{diag}}\left(\lambda_{1},\ldots,\lambda_{N}\right)$
is an $N\times N$ diagonal matrix whose $n$-th diagonal entry is
the corresponding eigenvalue of $\boldsymbol{u}_{n}$. Without loss
of generality, we assume that the values of the eigenvalues in $\boldsymbol{\Lambda}$
are arranged in descending order. i.e., $\lambda_{1}\geq\cdots\geq\lambda_{N}$.

Throughout this paper, we assume there is only one RF-chain in FAS
and thus only one port can be activated for communications. The received
signal of the $n$-th port is expressed as 
\begin{equation}
y_{n}=h_{n}x+w_{n},\;n=1,\ldots,N,\label{eq:3}
\end{equation}
where $h_{n}$ is the complex channel coefficient of the $n$-th port,
$x$ is the information signal with $\mathbb{E}\left[\left|x\right|^{2}\right]=P$
and $w_{n}$ is the additive white Gaussian noise of the $n$-th port
with zero mean and variance of $N_{0}$. Due to the spatial correlation
of the ports, $h_{n}$ can be modeled as 
\begin{equation}
h_{n}=\sum_{m=1}^{N}u_{n,m}\sqrt{\lambda_{m}}z_{m},\label{eq:4}
\end{equation}
where $u_{n,m}$ is the $(n,m)$-th entry of $\boldsymbol{U}$, $z_{m}=a_{m}+jb_{m}$,
where $a_{m},b_{m},\forall m,$ are independent and identically distributed
(i.i.d.) Gaussian random variables with zero mean and variance of
$\frac{1}{2}$. According to \cite{10103838}, (\ref{eq:4})
can also be approximated as 
\begin{equation}
\hat{h}_{n}=\Psi v_{n}+\sum_{m=1}^{\epsilon\text{-}{\rm rank}}u_{n,m}\sqrt{\lambda_{m}}z_{m},\label{eq:5}
\end{equation}
where $\epsilon\text{-}{\rm rank}$ is a modeling parameter, $\Psi=\sqrt{\sigma^{2}-\sum_{m=1}^{\epsilon\text{-}{\rm rank}}u_{n,m}^{2}\lambda_{m}}$,
$v_{n}=c_{n}+jd_{n}$ and $c_{n},d_{n},\forall n,$ are i.i.d. Gaussian
random variables with zero mean and variance of $\frac{1}{2}$.

To obtain the global optimum performance, FAS activates a port with
the maximum signal envelope \cite{9264694},\footnote{Due to the port spatial correlation, it is shown in \cite{Noor1}
that only a small number of observed ports/training is required to
obtain the full channel state information.} i.e., 
\begin{equation}
\left|h_{\text{FAS}}\right|=\max\left\{ \left|h_{1}\right|,\ldots,\left|h_{N}\right|\right\} .\label{eq:6}
\end{equation}
The instantaneous received SNR of the receiver is
found as 
\begin{align}
\Theta & =\left|h_{\text{{FAS}}}\right|^{2}\frac{P}{N_{0}}=\left|h_{\text{{FAS}}}\right|^{2}SNR,\label{eq:7}
\end{align}
where $SNR=\frac{P}{N_{0}}$ is the transmit SNR and its outage probability
is defined as 
\begin{align}
\mathcal{\mathbb{P}}\left\{ \log\left(1+\Theta\right)<q\right\} =\mathcal{\mathbb{P}}\left\{ \left|h_{\text{{FAS}}}\right|<\varOmega\right\} ,\label{eq:8}
\end{align}
where $\varOmega=\sqrt{\frac{2^{q}-1}{SNR}}$ and $q$ is the minimum
required rate. In addition, the diversity gain of FAS can be defined
as \cite{tse2005fundamentals} 
\begin{align}
\underset{SNR\rightarrow\infty}{\lim}-\frac{\log\mathcal{\mathbb{P}}_{e}\left(SNR\right)}{\log\left(SNR\right)}\overset{\left(a\right)}{=}\underset{SNR\rightarrow\infty}{\lim}-\frac{\log\mathcal{\mathbb{P}}\left\{ \log\left(1+\left|h_{\text{{FAS}}}\right|^{2}SNR\right)<q\right\} }{\log\left(SNR\right)}=D_{\rm{FAS}},\label{eq:9}
\end{align}
where $(a)$ follows from the fact that error probability and outage
probability differ by a constant shift at high SNR \cite{1221802}.

\section{Outage Probability and Diversity Gain of FAS}

As it is seen in (\ref{eq:4}), the complex channel coefficients $\boldsymbol{h}=\left[h_{1},\ldots,h_{N}\right]^{T}$
are correlated. Therefore, $\left|\boldsymbol{h}\right|$ is a correlated
Rayleigh random vector. We present the following lemmas to obtain
the closed-form outage probability and diversity gain of FAS.
\begin{lem}
\label{lem:PDF}The PDF of $\left|\boldsymbol{h}\right|$ can be approximated
as 
\begin{align}
 & f_{\left|\boldsymbol{h}\right|}\left(\left|h_{1}\right|,\ldots,\left|h_{N}\right|\right)\label{eq:10}\\
\approx~ & \eta\stackrel[s_{1}=0]{s_{0}}{\sum}\sum_{s_{2}=0}^{s_{1}}\ldots\sum_{s_{T}=0}^{s_{T-1}}\left(\frac{1}{2}\right)^{\sum_{t=1}^{T}s_{t}^{*}}\prod_{t=1}^{T}\beta\left(t,s_{t}^{*}\right)\sum_{\boldsymbol{v}\in\mathcal{V}}\left[\prod_{t=1}^{T}\left(\begin{array}{c}
s_{t}^{*}\\
v_{t}
\end{array}\right)\right]\left[\left(2\pi\right)^{N}\prod_{i=1}^{N}\boldsymbol{1}_{\left\{ \Delta_{i}=0\right\} }\right].\nonumber 
\end{align}
\end{lem}
\begin{IEEEproof}
\noindent See Appendix A. 
\end{IEEEproof}
In (\ref{eq:10}), $\eta=\frac{\stackrel[n=1]{N}{\prod}\left|h_{n}\right|}{\pi^{N}\text{{det}\ensuremath{\left(\boldsymbol{J}\right)}}}\exp\left\{ -\frac{\sum_{n=1}^{N}\left|h_{n}\right|^{2}K_{n,n}}{\text{{det}\ensuremath{\left(\boldsymbol{J}\right)}}}\right\} $,
$T=\frac{N\left(N-1\right)}{2}$ , $\beta\left(t,s_{t}^{*}\right)\triangleq\frac{\zeta_{t}^{s_{t}^{*}}}{s_{t}^{*}!}$,
$\zeta_{t}=-\frac{2K_{m,n}\left|h_{n}\right|\left|h_{m}\right|}{\text{{det}\ensuremath{\left(\boldsymbol{J}\right)}}}$
and $s_{t}^{*}=s_{t}-s_{t+1}$ with $s_{T+1}=0$. Throughout this
paper, the subscript $t$ and $m,n$ are related as follows: $t=n+\left(m-1\right)N-\frac{m\left(m+1\right)}{2}$,
$m<n,$ while $m,n$ can be obtained from $t$ with $m=\min\:m'\in\mathbb{Z}$
subject to $\sum_{i=1}^{m'}\left(N-i\right)>t$ and $n=t-\left(m-1\right)N+\frac{m\left(m+1\right)}{2}$.

Note that $s_{0}$ is a finite constant which has to be large for
the approximation to be accurate. In addition, $\boldsymbol{v}=\left[v_{1},\ldots,v_{T}\right]^{T}$,
$\mathcal{V}$ denotes the set of all the possible permutations and
$\Delta_{i}=\sum_{n=1}^{N}G_{i,n}-\sum_{n=1}^{N}G_{n,i}-G_{i,i}$.
Furthermore, $K_{m,n}$ is the $(m,n)$-th entry of $\boldsymbol{K}$
where $\boldsymbol{K}$ is the co-factor of $\boldsymbol{J}$, and
$G_{m,n}$ is the $(m,n)$-th entry of $\boldsymbol{G}$ where $\boldsymbol{G}$
is defined as 
\begin{equation}
\boldsymbol{G}=\left[\begin{array}{ccccc}
0 & \gamma_{1} & \gamma_{2} & \cdots & \gamma_{N-1}\\
 &  & \gamma_{N} & \cdots & \gamma_{2N-3}\\
\vdots &  & \ddots &  & \vdots\\
 &  &  &  & \gamma_{T}\\
0 &  & \cdots &  & 0
\end{array}\right],\label{eq:11}
\end{equation}
and $\gamma_{t}=2v_{t}-s_{t}^{*}\in\mathbb{Z}$.
\begin{lem}
\label{lem:CDF}The CDF of $\left|\boldsymbol{h}\right|$ can be approximated
as 
\begin{align}
F_{\left|\boldsymbol{h}\right|}\left(R_{1},\ldots,R_{N}\right) & \approx\stackrel[s_{1}=0]{s_{0}}{\sum}\sum_{s_{2}=0}^{s_{1}}\ldots\sum_{s_{T}=0}^{s_{T-1}}\frac{g\left(\boldsymbol{s}^{*}\right)}{\pi^{N}\text{{\rm {det}}\ensuremath{\left(\boldsymbol{J}\right)}}}\prod_{t=1}^{T}\frac{\left(-K_{m,n}\right)^{s_{t}^{*}}}{s_{t}^{*}!\text{{\rm {det}}\ensuremath{\left(\boldsymbol{J}\right)}}^{s_{t}^{*}}}\times\label{eq:12}\\
 & \stackrel[n=1]{N}{\prod}\frac{1}{2}\left(\frac{\text{\ensuremath{K_{n,n}}}}{\text{{\rm {det}}\ensuremath{\left(\boldsymbol{J}\right)}}}\right)^{-\frac{\bar{s}_{n}}{2}-\frac{1}{2}}\left[\Gamma\left(\frac{1+\bar{s}_{n}}{2}\right)-\Gamma\left(\frac{1+\bar{s}_{n}}{2},\frac{\text{\ensuremath{K_{n,n}}}R_{n}^{2}}{\text{{\rm {det}}\ensuremath{\left(\boldsymbol{J}\right)}}}\right)\right].\nonumber 
\end{align}
\end{lem}
\begin{IEEEproof}
See Appendix B. 
\end{IEEEproof}
In (\ref{eq:12}), we have $\bar{s}_{n}=\sum_{i=1}^{N}S_{n,i}^{*}+\sum_{i=1}^{n-1}S_{i,n}^{*}+1$
where $S_{i,n}^{*}$ is the $\left(i,n\right)$-th entry of $\boldsymbol{S}^{*}$
and $\boldsymbol{S}^{*}$ is introduced in (\ref{eq:B6}). Furthermore,
\begin{equation}
g\left(\boldsymbol{s}^{*}\right)=\left(\frac{1}{2}\right)^{\sum_{t=1}^{T}s_{t}^{*}}\sum_{\boldsymbol{v}\in\mathcal{V}}\left[\prod_{t=1}^{T}\left(\begin{array}{c}
s_{t}^{*}\\
v_{t}
\end{array}\right)\right]\left(2\pi\right)^{N}\prod_{i=1}^{N}\boldsymbol{1}_{\left\{ \Delta_{i}=0\right\} }.\label{eq:13}
\end{equation}
The expressions in (\ref{eq:10}) and (\ref{eq:12}) are extremely
complicated. Nevertheless, they enable us to obtain more insightful
derivations as shown later in this paper. Using the above lemmas,
we present the following theorems.
\begin{thm}
\label{thm:outage}The outage probability of FAS can be approximated
in a closed-form expression as 
\begin{align}
\mathcal{\mathbb{P}}\left\{ \left|h_{\text{{FAS}}}\right|<\varOmega\right\}  & =F_{\left|\boldsymbol{h}\right|}\left(\varOmega,\ldots,\varOmega\right)\label{eq:14}\\
 & \approx\stackrel[s_{1}=0]{s_{0}}{\sum}\sum_{s_{2}=0}^{s_{1}}\ldots\sum_{s_{T}=0}^{s_{T-1}}\frac{g\left(\boldsymbol{s}^{*}\right)}{\pi^{N}\text{{\rm {det}}\ensuremath{\left(\boldsymbol{J}\right)}}}\prod_{t=1}^{T}\frac{\left(-K_{m,n}\right)^{s_{t}^{*}}}{s_{t}^{*}!\text{{\rm {det}}\ensuremath{\left(\boldsymbol{J}\right)}}^{s_{t}^{*}}}\times\nonumber \\
 & \stackrel[n=1]{N}{\prod}\frac{1}{2}\left(\frac{\text{\ensuremath{K_{n,n}}}}{\text{{\rm {det}}\ensuremath{\left(\boldsymbol{J}\right)}}}\right)^{-\frac{\bar{s}_{n}}{2}-\frac{1}{2}}\left[\Gamma\left(\frac{1+\bar{s}_{n}}{2}\right)-\Gamma\left(\frac{1+\bar{s}_{n}}{2},\frac{\text{\ensuremath{K_{n,n}}}\varOmega^{2}}{\text{{\rm {det}}\ensuremath{\left(\boldsymbol{J}\right)}}}\right)\right].\nonumber 
\end{align}
\end{thm}
\begin{IEEEproof}
The result can be obtained using Lemma \ref{lem:CDF} and substituting
$R_{1}=\cdots=R_{N}=\varOmega$. 
\end{IEEEproof}
\begin{rem}
According to \cite{10103838}, $\boldsymbol{h}$ can be modeled
using $\hat{\boldsymbol{h}}=\left[\hat{h}_{1},\ldots,\hat{h}_{N}\right]^{T}$
and using the latter model, they show that the outage probability
of FAS can be approximated by 
\begin{align}
F_{\left|h_{\text{FAS}}\right|}\left(\varOmega\right)\approx & \Bigg[\prod_{n=1}^{N}\intop_{0}^{\infty}\frac{1}{\sum_{m=1}^{\epsilon\text{-}{\rm rank}}u_{n,m}^{2}\lambda_{m}}\exp\left(-\frac{r}{\sum_{m=1}^{\epsilon\text{-}{\rm rank}}u_{n,m}^{2}\lambda_{m}}\right)\left(1-Q_{1}\left(\frac{\sqrt{2r}}{\Psi},\frac{\sqrt{2}\varOmega}{\Psi}\right)\right)^{L}dr\Bigg]^{\frac{1}{L}},\label{eq:15}
\end{align}
where $Q_{1}\left(\cdot,\cdot\right)$ is the Marcum-Q function and
$L=\min\left\{ \frac{1.52\left(N-1\right)}{2\pi W},N\right\} $. Note
that (\ref{eq:15}) is a remarkable expression as each $n$ term only
has a single integral. Nevertheless, we found that it is challenging
to obtain deeper insights from this expression. 
\end{rem}
\begin{thm}
\label{thm:outage_highSNR} The outage probability of FAS at high
SNR is given by 
\begin{equation}
\mathcal{\mathbb{P}}\left\{ \left|h_{\text{{FAS}}}\right|<\varOmega\right\} =\frac{1}{\text{{\rm {det}}\ensuremath{\left(\boldsymbol{J}\right)}}}\varOmega^{2N}+o\left(\frac{1}{SNR^{N}}\right).\label{eq:16}
\end{equation}
\end{thm}
\begin{IEEEproof}
See Appendix C. 
\end{IEEEproof}
\begin{thm}
\label{thm:diversity_gain}The diversity gain of FAS is approximately
expressed as 
\begin{equation}
D_{\text{FAS}}\approx\min\left\{ N,N'\right\} ,\label{eq:17}
\end{equation}
where $N'$ is the numerical rank of $\boldsymbol{J}'$ such that
$\boldsymbol{J}'$ is the covariance matrix as defined in (\ref{eq:2})
with $N\rightarrow\infty$ for a fixed $W$. 
\end{thm}
\begin{IEEEproof}
See Appendix D. 
\end{IEEEproof}
In Theorem \ref{thm:outage_highSNR}, we can interpret $\text{{det}}\left(\boldsymbol{J}^{-1}\right)$
as the penalty term and $\varOmega^{2}$ as gain of FAS that scales
exponentially w.r.t. $N$. Meanwhile, the term with little-o can be
ignored as it approaches zero if the $SNR$ is high. Nevertheless,
in Theorem \ref{thm:diversity_gain}, we can see that the diversity
gain is limited by $\min\left\{ N,N'\right\} $. Thus, increasing
$N$ over $N'$ might not be useful. Notice that these interpretations
cannot be directly obtained from (\ref{eq:15}).

\section{Suboptimal Solution: FAS with $N^{*}$ Ports}

At a fundamental level, \cite{4065786} showed that increasing the
number of channels (or ports) would yield a diminishing gain (i.e.,
the average received SNR gain is $\sum_{n}^{N}\frac{1}{n}$.). In
fact, \cite{10103838} showed that for a fixed $W$, the outage
probability of FAS might remain similar after some $N$. For ease
of expositions, we denote this $N$ as $N^{*}$ where $N^{*}\leq N$.

To the best of our knowledge, little is known about $N^{*}$. In fact,
it is very challenging to obtain $N^{*}$ as it varies with the parameter
$W$ or more precisely the correlation matrix $\boldsymbol{J}$.\footnote{Referring to (\ref{eq:1}) and (\ref{eq:2}), we can see that $N^{*}$
depends on the parameter $W$.} Yet, finding $N^{*}$ is essential in both theory and practice since
it helps FAS to achieve an efficient performance with a minimal number
of ports. In this section, we present a simple method to approximate
$N^{*}$ for a given $W$.

To begin with, we present the following theorem.
\begin{thm}
\label{thm:approx}Suppose the channels of FAS with $N$ ports are
denoted by $\boldsymbol{h}$. Then $\boldsymbol{h}$ can be well-approximated
by $\tilde{\boldsymbol{h}}=\left[\tilde{h}_{1},\ldots,\tilde{h}_{N}\right]^{T}$
where 
\begin{equation}
\tilde{h}_{n}=\sum_{m=1}^{\tilde{N}}u_{n,m}\sqrt{\lambda_{m}}z_{m},\label{eq:18}
\end{equation}
where $\tilde{N}$ is the numerical rank of $\boldsymbol{J}$. That
is, the PDF and CDF of $\boldsymbol{h}$ and $\tilde{\boldsymbol{h}}$
are similar. 
\end{thm}
\begin{IEEEproof}
Let $\tilde{N}$ be the numerical rank of $\boldsymbol{J}$ where
$\tilde{N}\leq N$. Using the definition of numerical rank, we have
$\lambda_{n}<\epsilon$ for $n\in\left\{ \tilde{N}+1,\ldots,N\right\} $
where $\epsilon\approx0$. According to Eckart-Young-Mirsky theorem
\cite{GOLUB1987317}, the optimal $\tilde{\boldsymbol{J}}$ that minimizes
the Frobenius norm between matrix $\boldsymbol{J}$ and $\tilde{\boldsymbol{J}}$
subject to the constraint that rank$\left(\tilde{\boldsymbol{J}}\right)\leq\tilde{N}$
is $\tilde{\boldsymbol{J}}=\boldsymbol{U}\tilde{\boldsymbol{\Lambda}}\boldsymbol{U}^{H}$
where $\tilde{\boldsymbol{\Lambda}}=\text{{diag}}\left(\lambda_{1},\ldots,\lambda_{\tilde{N}},0,\ldots0\right)$.

Using this insight, we introduce $\tilde{\boldsymbol{h}}$ as defined
in Theorem \ref{thm:approx} where the covariance of $\tilde{\boldsymbol{h}}$
is $\tilde{\boldsymbol{J}}$ (i.e., the best approximation of $\boldsymbol{J}$
for rank$\left(\tilde{\boldsymbol{J}}\right)\leq\tilde{N}$). As a
result, we can well-approximate $\boldsymbol{h}$ using $\tilde{\boldsymbol{h}}$
since the Frechet distance between the two distributions is \cite{DOWSON1982450}
\begin{align}
W_{2}\left(\mathcal{{CN}}\left(0_{N\times1},\boldsymbol{J}\right),\mathcal{CN}\left(0_{N\times1},\tilde{\boldsymbol{J}}\right)\right) & =\text{\ensuremath{\left\Vert \left(\boldsymbol{\Lambda}\right)^{\frac{1}{2}}-\left(\tilde{\boldsymbol{\Lambda}}\right)^{\frac{1}{2}}\right\Vert }}_{F}^{2}\label{eq:19}\\
 & \approx0.\nonumber 
\end{align}
\end{IEEEproof}
\begin{cor}
If we have the exact eigenvalues and rank of $\boldsymbol{J}$, then
$\boldsymbol{h}=\tilde{\boldsymbol{h}}$. 
\end{cor}
\begin{IEEEproof}
Let $\boldsymbol{\Lambda}$ and $\tilde{N}$ be the exact eigenvalues
and rank of $\boldsymbol{J}$. Using the definition of rank, we have
$\lambda_{n}=0$ for $n\in\left\{ \tilde{N}+1,\ldots,N\right\} $.
It then follows that the Frechet distance between the distributions
of $\boldsymbol{h}$ and $\tilde{\boldsymbol{h}}$ is zero. 
\end{IEEEproof}
As seen in (\ref{eq:19}), it is the eigenvalues of correlation matrix
that play a critical role in the channel approximation. Motivated
by this insight, we introduce a new formula as follows: 
\begin{align}
\varepsilon_{N^{*}} & =S_{N}-S_{N^{*}}\label{eq:20}\\
 & =\sigma^{2}-S_{N^{*}},\nonumber 
\end{align}
where $S_{N^{*}}=\frac{1}{N}\sum_{n=1}^{N^{*}}\lambda_{n}$. Note
that (\ref{eq:20}) is analogous to (\ref{eq:19}) in the sense that
the left hand side of (\ref{eq:20}) measures the gap between the
distributions of $\boldsymbol{h}$ and $\boldsymbol{h}^{*}$, where
$\boldsymbol{h}^{*}$ is similarly defined as in (\ref{eq:18}) but
we instead replace $\tilde{N}$ with $N^{*}$ and impose that $N^{*}\leq\tilde{N}$.
Meanwhile, on the right hand side of (\ref{eq:20}), we consider the
average eigenvalues of $\boldsymbol{J}^{*}$, where $\boldsymbol{J}^{*}$
is the covariance of $\boldsymbol{h}^{*}$.

To reduce the number of required ports, we define $\varepsilon_{\text{tol}}>0$
and find the smallest integer $N^{*}$ such that $\varepsilon_{\text{tol}}\geq\varepsilon_{N^{*}}$.
Since $\boldsymbol{J}^{*}$ only has $N^{*}$ dominant eigenvalues,
we propose to employ a suboptimal FAS with $N^{*}$ ports. Interestingly,
$\varepsilon_{\text{tol}}$ has a nice heuristic interpretation in
practice. Specifically, it defines the sub-optimality of the proposed
FAS, i.e., the proposed FAS is near optimal if $\varepsilon_{\text{tol}}$
is small and less optimal if $\varepsilon_{\text{tol}}$ is large.

By fixing $\varepsilon_{\text{tol}}$ appropriately,\footnote{We recommend to set $\varepsilon_{\text{tol}}=0.01\sigma^{2}$ (i.e.,
the average eigenvalues of $\boldsymbol{J}^{*}$ is $99\%$ of that
of $\boldsymbol{J}$)} we observe that FAS with $N^{*}$ ports yields considerable improvement
over all FAS with $N<N^{*}$ ports while most of the FAS with $N>N^{*}$
ports yields marginal improvement over FAS with $N-1$ ports. Note
that we usually have $N^{*}<\tilde{N}$ if $\boldsymbol{J}$ is ill-conditioned
and $N^{*}=\tilde{N}$ if $\boldsymbol{J}$ is well-conditioned.

The method of approximating $N^{*}$ is given in Algorithm 1. To measure
the computational complexity of our algorithm, we consider the floating-point
operations (flops). A flop is defined as one addition, subtraction,
multiplication or division of two floating point numbers \cite{boyd2004convex}.
In Algorithm 1, computing $\boldsymbol{J}$ and $\boldsymbol{U}\boldsymbol{\Lambda}\boldsymbol{U}^{H}$
requires $6N^{2}$ and $21N^{3}$ flops, respectively \cite{ford2014numerical}.
Computing $\varepsilon_{n}$ requires $n+1$ flops for each $n$.
Therefore, the total flops of Algorithm 1 is $21N^{3}+6N^{2}+\frac{1}{2}N^{*2}+\frac{3}{2}N^{*}$,
which has a polynomial time-complexity of $\mathcal{{O}}\left(N^{3}\right)$
since $N^{*}\leq N$. In other words, Algorithm 1 is only dominated
by the computation of $\boldsymbol{U}\boldsymbol{\Lambda}\boldsymbol{U}^{H}$.

Note that $N^{*}$ is also useful in theory. For example, Lemma \ref{lem:PDF}
and \ref{lem:CDF} and Theorem \ref{thm:outage}, \ref{thm:outage_highSNR},
and \ref{thm:diversity_gain} are incalculable if $\boldsymbol{J}$
is near-singular. To address this, we present the following theorem. 
\begin{thm}
\label{thm:singular} If $\boldsymbol{J}$ is near-singular, then
we can approximate the channels of FAS with $N$ ports using $N^{*}$
ports from a computational perspective. Nevertheless, a small gap
between the channel distributions of FAS with $N$ ports and that
of $N^{*}$ ports might exist. 
\end{thm}
\begin{IEEEproof}
If $\boldsymbol{J}$ is near-singular, then one or more entries are
almost linear combinations of the other entries. Thus, we can remove
these nearly-dependent entries and only consider $N^{*}$ independent
entries. Since FAS with $N^{*}$ ports has $N^{*}$ dominant eigenvalues,
Lemma \ref{lem:PDF} and \ref{lem:CDF} and Theorem \ref{thm:outage},
\ref{thm:outage_highSNR}, and \ref{thm:diversity_gain} are calculable.
Nevertheless, there might be a small gap between the channel distributions
of FAS with $N$ ports and that of $N^{*}$ ports since the entries
are nearly-dependent only. 
\end{IEEEproof}
\begin{algorithm}[t]
\begin{algorithmic}[1]

\STATE \textbf{Input}: $W,\varepsilon_{\text{tol}}$; \textbf{Output}:
$N^{*}$

\STATE Compute $\boldsymbol{J}=\boldsymbol{U}\boldsymbol{\Lambda}\boldsymbol{U}^{H}$

\STATE Define $n=1$ and compute $\varepsilon_{n}$

\STATE\textbf{ While} $\varepsilon_{\text{tol}}<\varepsilon_{n}$
and $n<\tilde{N}$

\STATE $\qquad n=n+1$

\STATE $\qquad\varepsilon_{n}=\sigma^{2}-S_{n}$

\STATE \textbf{end}

\STATE Return $n$ as $N^{*}$

\end{algorithmic}

\caption{\label{alg:algo1}Method of approximating $N^{*}$ given $W$}
\end{algorithm}

\section{Results and Discussions}

In this section, we present simulation results to better understand
the performance of FAS. We focus on the design of an efficient FAS
as well as the factors that limit its performance. Unless stated otherwise,
we assume that $\sigma^{2}=1$, $N=50$, $W=0.5$, $q=10$ and $SNR=30$dB.

\begin{figure}
\centering{}\subfloat[]{\begin{centering}
\includegraphics[scale=0.52]{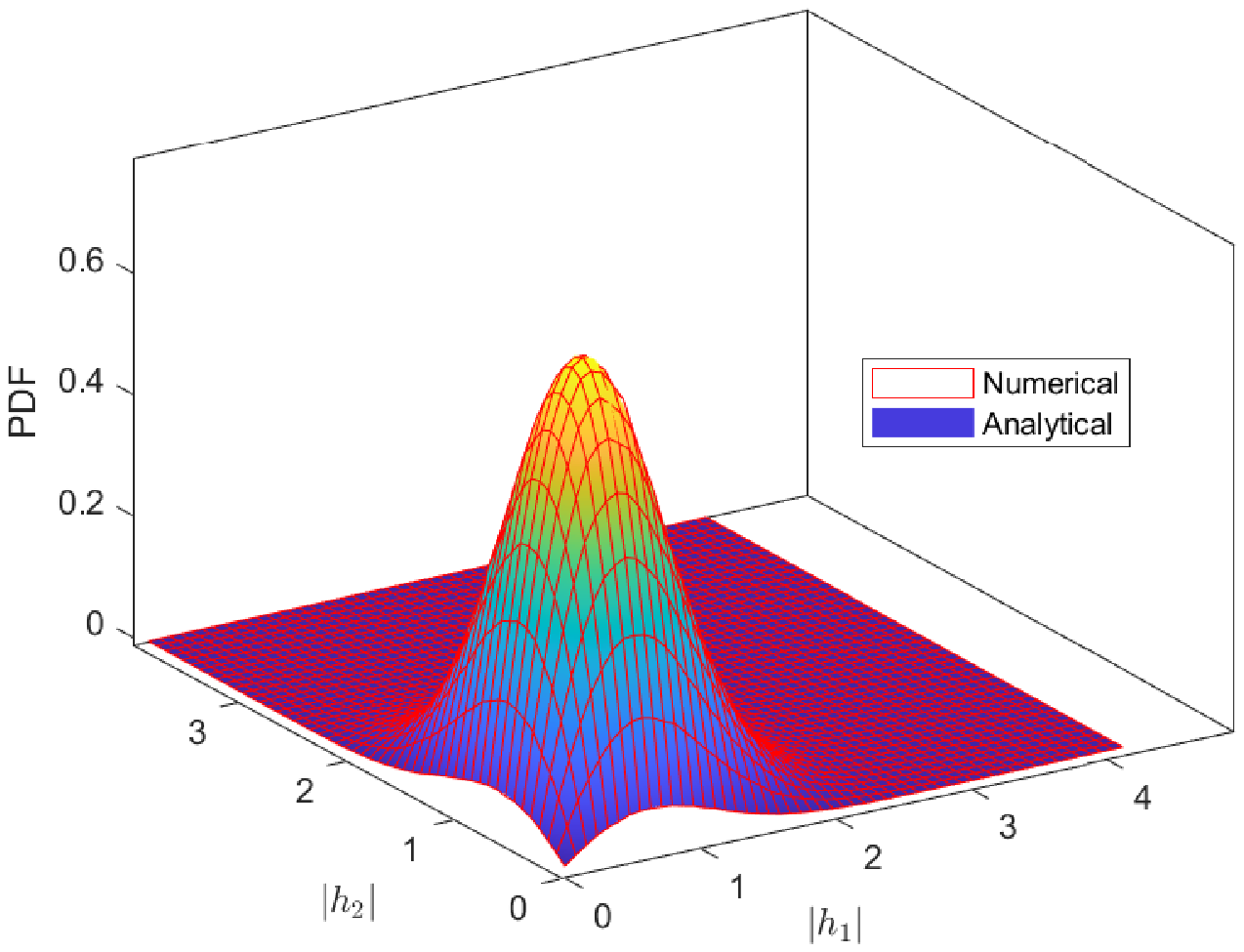} 
\par\end{centering}
}\hfill{}\subfloat[]{\begin{centering}
\includegraphics[scale=0.52]{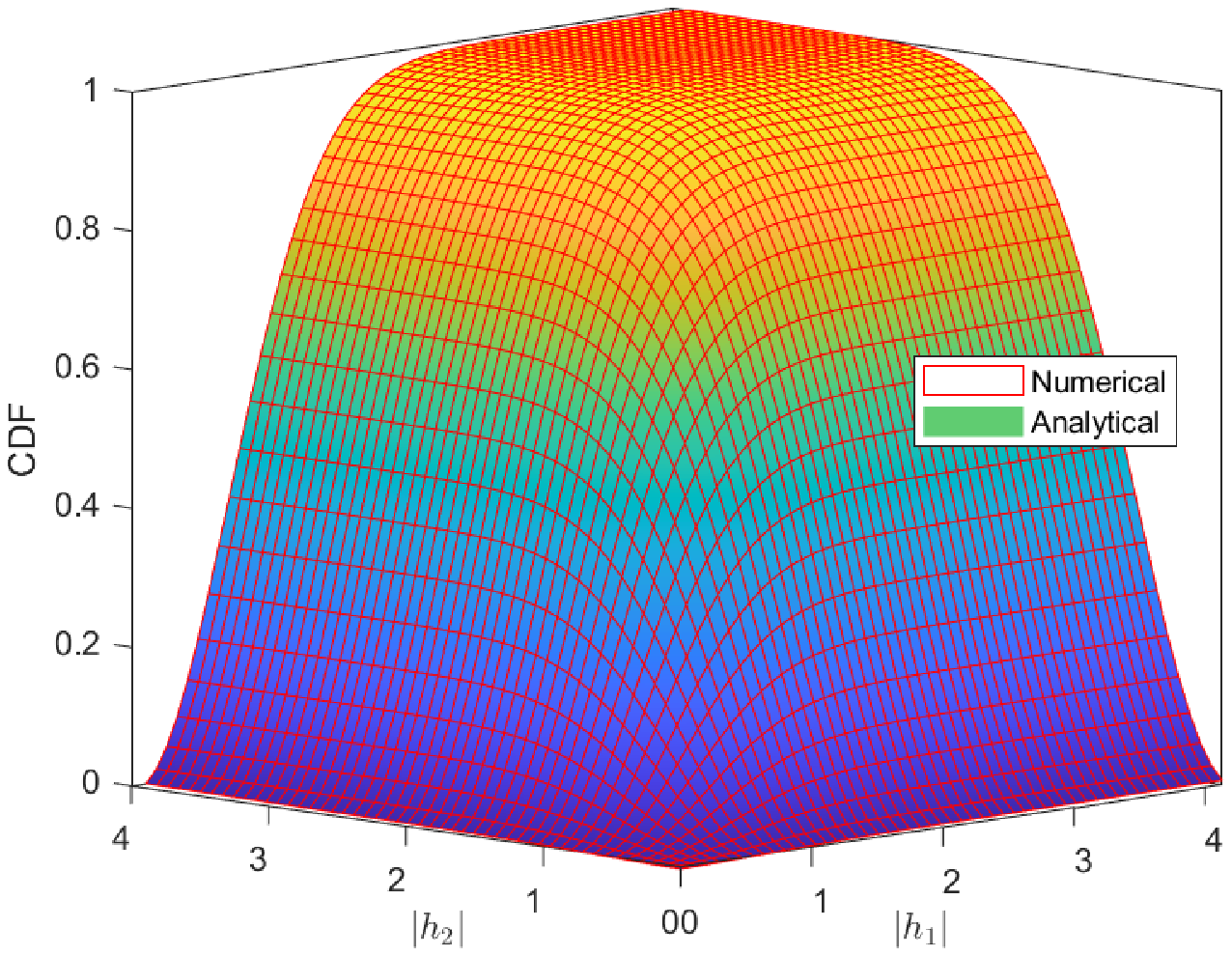} 
\par\end{centering}
\centering{}

}\caption{\label{fig:fig1}FAS with 2 ports: a) joint PDF; b) joint CDF.}
\end{figure}

Firstly, we demonstrate the accuracy of (\ref{eq:10}) and (\ref{eq:12}).
In order to visualize the joint PDF and CDF of $\left|\boldsymbol{h}\right|$,
we consider a FAS with 2 ports (i.e., $N=2$). In Fig. \ref{fig:fig1},
the red grid represents the numerical PDF/CDF while the solid surface
is the analytical PDF/CDF. As observed, the approximation of the PDF/CDF
of $\left|\boldsymbol{h}\right|$ matches closely with the numerical
ones over all the distributed region. Still, it is worth noting that
(\ref{eq:10}) and (\ref{eq:12}) are very complicated. Thus, approximations
with simpler expressions remain desirable. 
\begin{figure}
\begin{centering}
\includegraphics[scale=0.52]{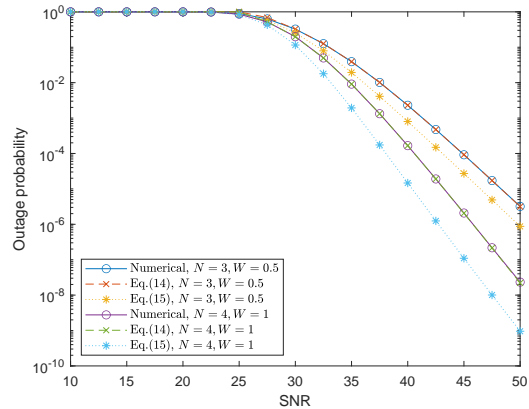} 
\par\end{centering}
\caption{\label{fig:fig2}Outage probability of FAS versus SNR.}
\end{figure}

Fig. \ref{fig:fig2} compares the outage probability
of FAS to (\ref{eq:14}) and (\ref{eq:15}). As observed, (\ref{eq:14})
is more accurate because the analytical expression is derived directly
from the multivariate correlated Rayleigh distributions and the approximation
is only used when truncating the infinite series to a finite one.
Here, we assume that $s_{0}=20$. Compared to the numerical result,
the truncation error is negligible as long as $s_{0}$ is sufficiently
large. In contrast, (\ref{eq:15}) is less accurate because the outage
probability of FAS is approximated using the power of single integrals
where such simplification may lead to some inaccuracies. Nevertheless,
it is worth highlighting that (\ref{eq:14}) can only be computed
for small $N$ as its expression is highly complicated. Thus, (\ref{eq:15})
is still useful for large $N$.
\begin{figure}
\begin{centering}
\subfloat[]{\begin{centering}
\includegraphics[scale=0.52]{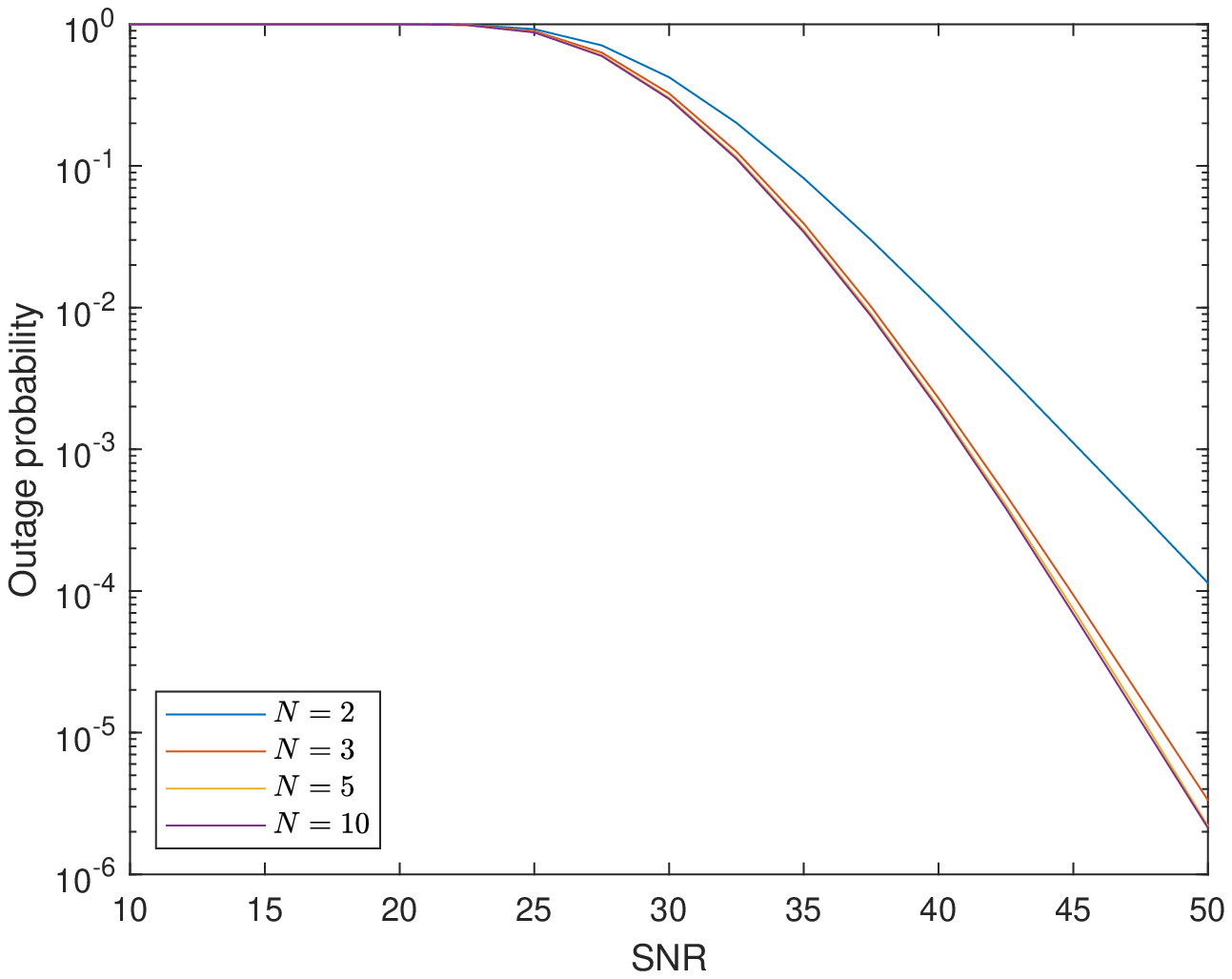} 
\par\end{centering}
\centering{}

}\hfill{}\subfloat[]{\begin{centering}
\includegraphics[scale=0.52]{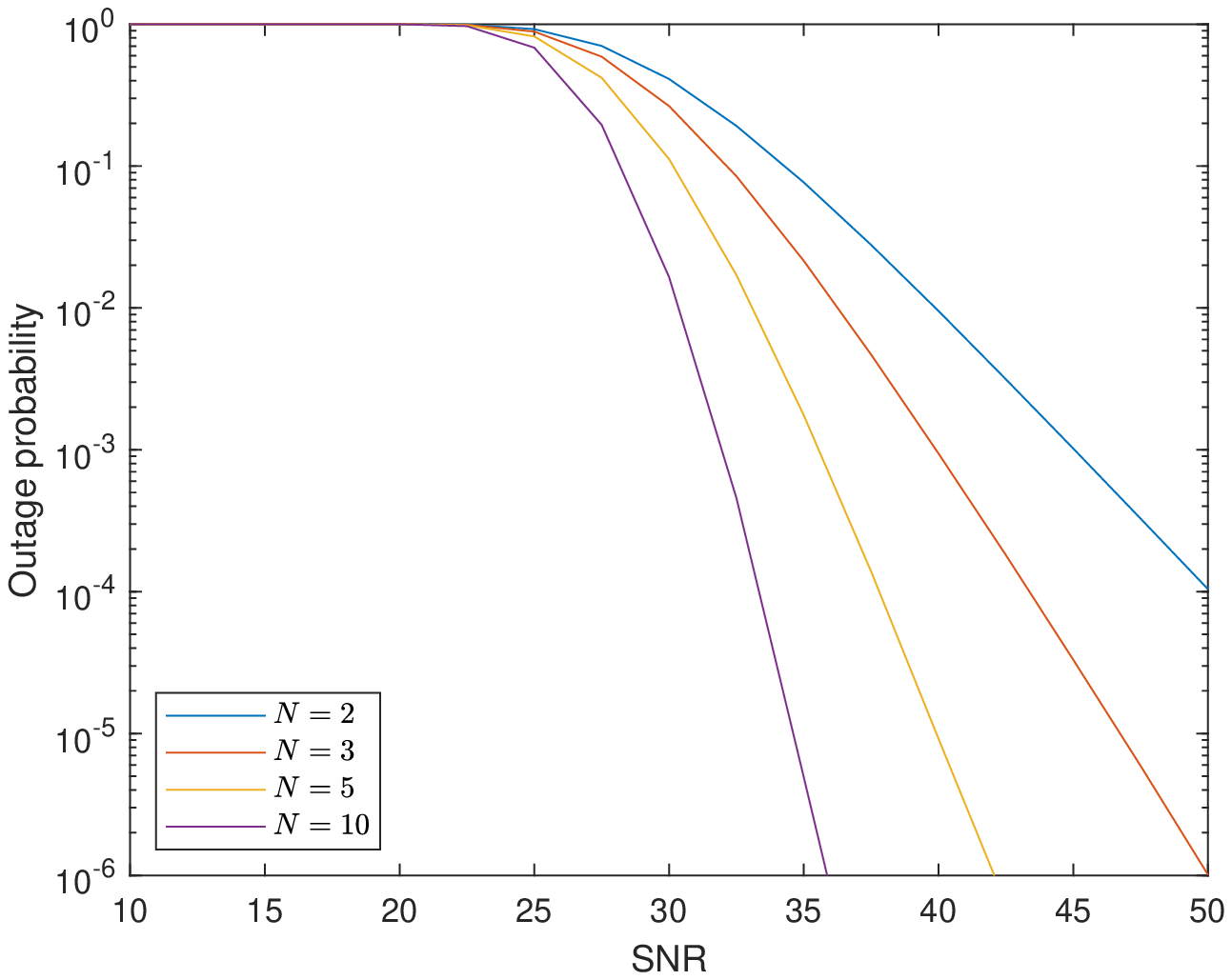} 
\par\end{centering}
}
\par\end{centering}
\caption{\label{fig:fig3}Outage probability of FAS versus SNR for different
$N$ and $W$: a) $W=0.5$; b) $W=10$.}
\end{figure}

In Fig. \ref{fig:fig3}, we compute the outage probability of FAS
versus SNR for different $N$ and $W$. Comparing Fig. \ref{fig:fig3}(a)
and Fig. \ref{fig:fig3}(b), we can clearly see that the outage probability
is mainly limited by $W$. In particular, if $W$ is small and $N$
is large, the outage probability remains similar which is in alignment
with the findings of \cite{10103838}. Nevertheless, if $W$
is sufficiently large, the outage probability decreases significantly
as $N$ increases.

\begin{figure}
\begin{centering}
\includegraphics[scale=0.52]{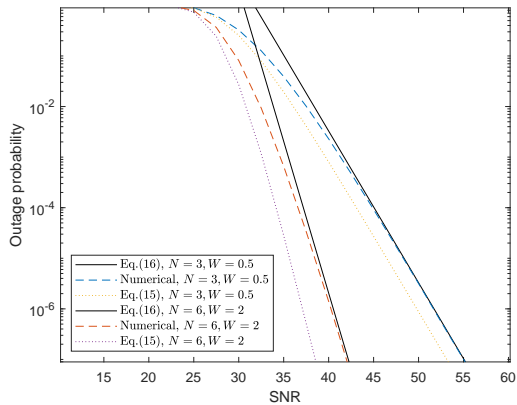} 
\par\end{centering}
\caption{\label{fig:fig4}Outage probability of FAS at high SNR.}
\end{figure}

To better understand this, we further compare the outage probability of FAS to (\ref{eq:15}) and (\ref{eq:16}) in Fig. \ref{fig:fig4}. Compared to the numerical result, we can see that (\ref{eq:15}) is less accurate while (\ref{eq:16}) is accurate as $SNR$ increases. Specifically, (\ref{eq:16}) is much more accurate as $SNR$ increases because we apply Taylor series approximation at around zero which corresponds to asymptotically high SNR. Hence, the error becomes negligible at high SNR. From (\ref{eq:16}), we learn that $\text{{det}}\left(\boldsymbol{J}^{-1}\right)$ plays a critical role in the performance of FAS. In particular, $\boldsymbol{J}$ has to be well-conditioned in order for $\varOmega^{2N}$ to be the dominant term. If $\boldsymbol{J}$ is near-singular, then $N$ is no longer important. This is because $\text{{det}}\left(\boldsymbol{J}^{-1}\right)$ cannot be compensated by $\varOmega^{2N}$. To make $\boldsymbol{J}$ a well-conditioned matrix, we can either increase $W$ for a fixed $N$ or decrease $N$ for a fixed $W$. Nevertheless, we believe that larger $N$ does not cause any harm to the system in practice. It only makes the theoretical analysis harder.

\begin{figure}
\begin{centering}
\subfloat[]{\begin{centering}
\includegraphics[scale=0.52]{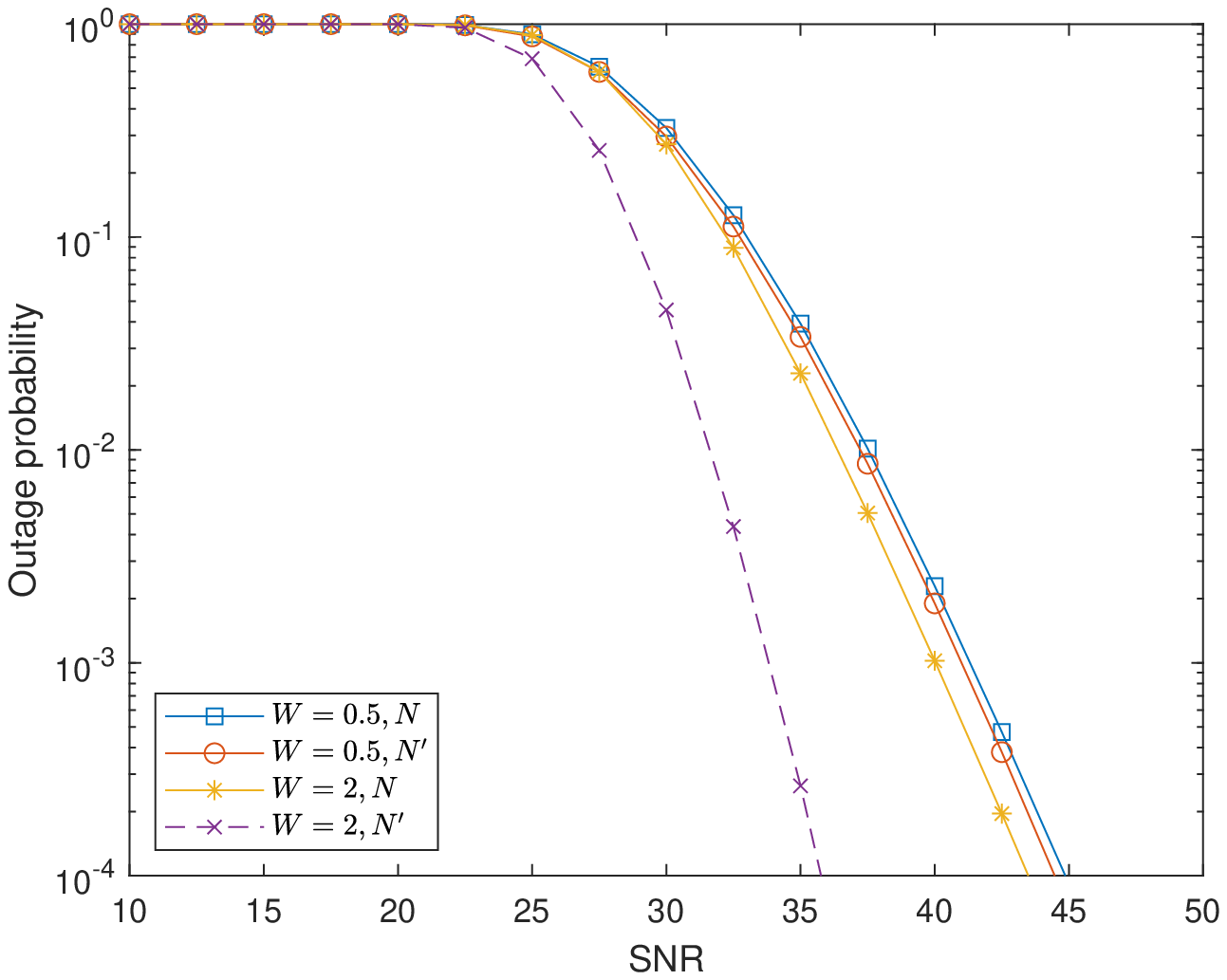} 
\par\end{centering}
\centering{}

}\hfill{}\subfloat[]{\begin{centering}
\includegraphics[scale=0.52]{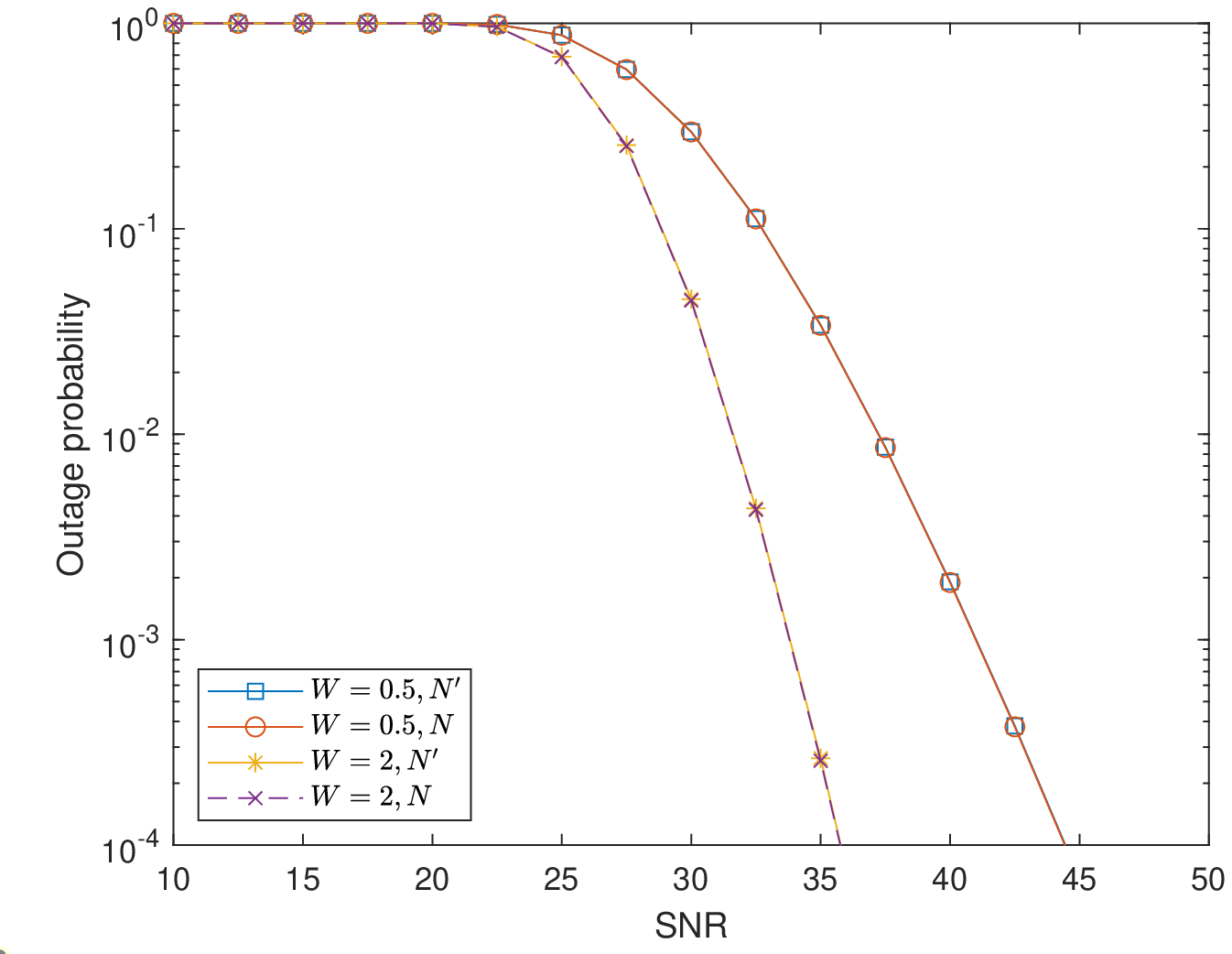} 
\par\end{centering}
}
\par\end{centering}
\caption{\label{fig:fig5}Outage probability of FAS with $N$ ports versus
$N'$ ports: a) $N=3<N'$; b) $N=50>N'$ .}
\end{figure}

As shown in Fig. \ref{fig:fig5}(a), we compare the outage probability
of FAS with $N$ ports and that of $N'$ ports for different $W$
where $N<N'$. As it is seen, the outage probability of the earlier
is lower bounded by the latter regardless of $W$. In Fig. \ref{fig:fig5}(b),
we investigate the opposite case where $N>N'$. As observed, the outage
probability of FAS with $N$ ports and that of $N'$ ports are the
same for different $W$. Thus, the diversity gain of FAS is limited
by $\min\left\{ N,N'\right\} $, which verifies Theorem \ref{thm:diversity_gain}.
Theorem \ref{thm:diversity_gain} also suggests that increasing the
ports beyond $N'$ provides no improvement in a point-to-point setting.

\begin{figure}
\begin{centering}
\subfloat[]{\begin{centering}
\includegraphics[scale=0.52]{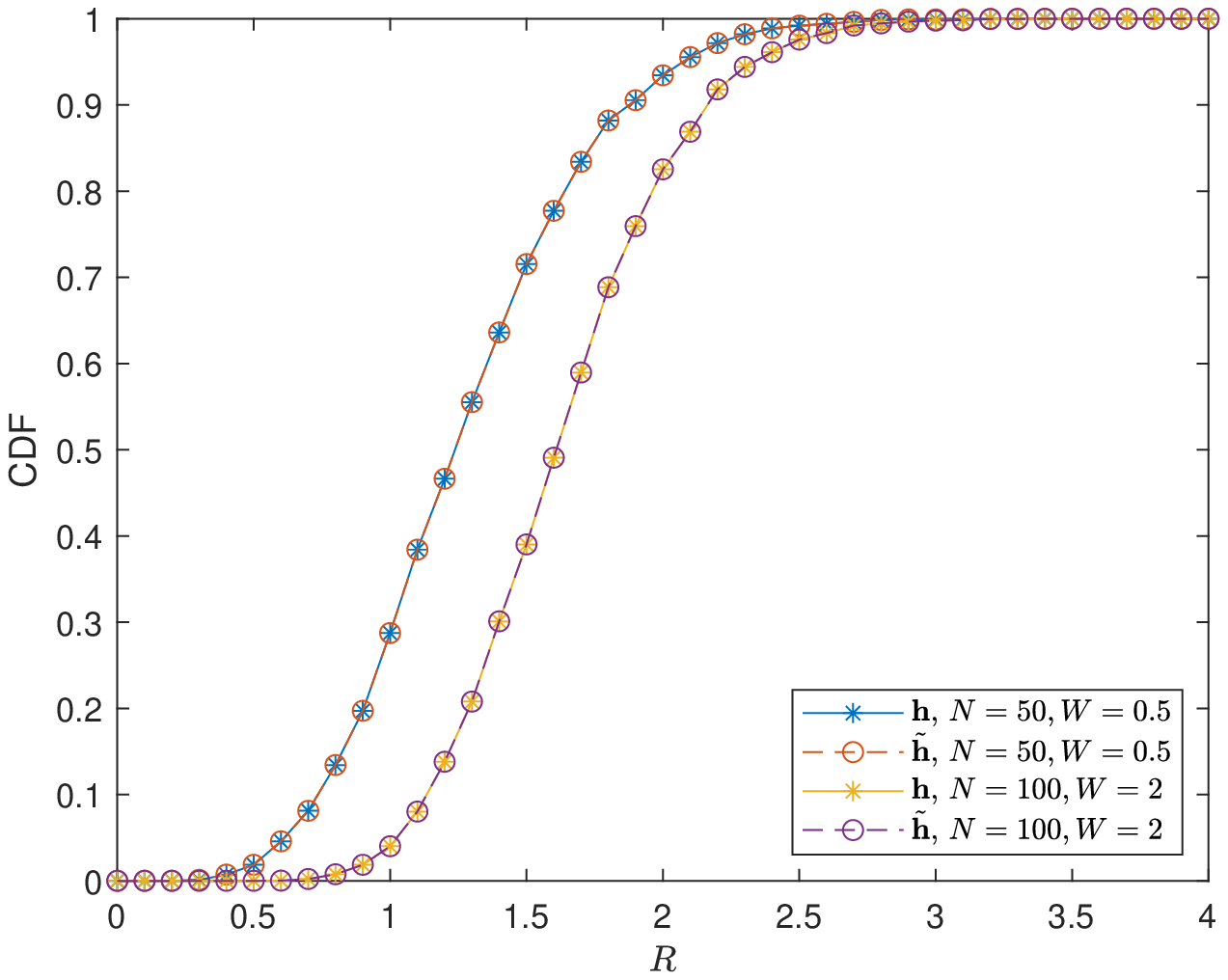} 
\par\end{centering}
\centering{}

}\hfill{}\subfloat[]{\begin{centering}
\includegraphics[scale=0.52]{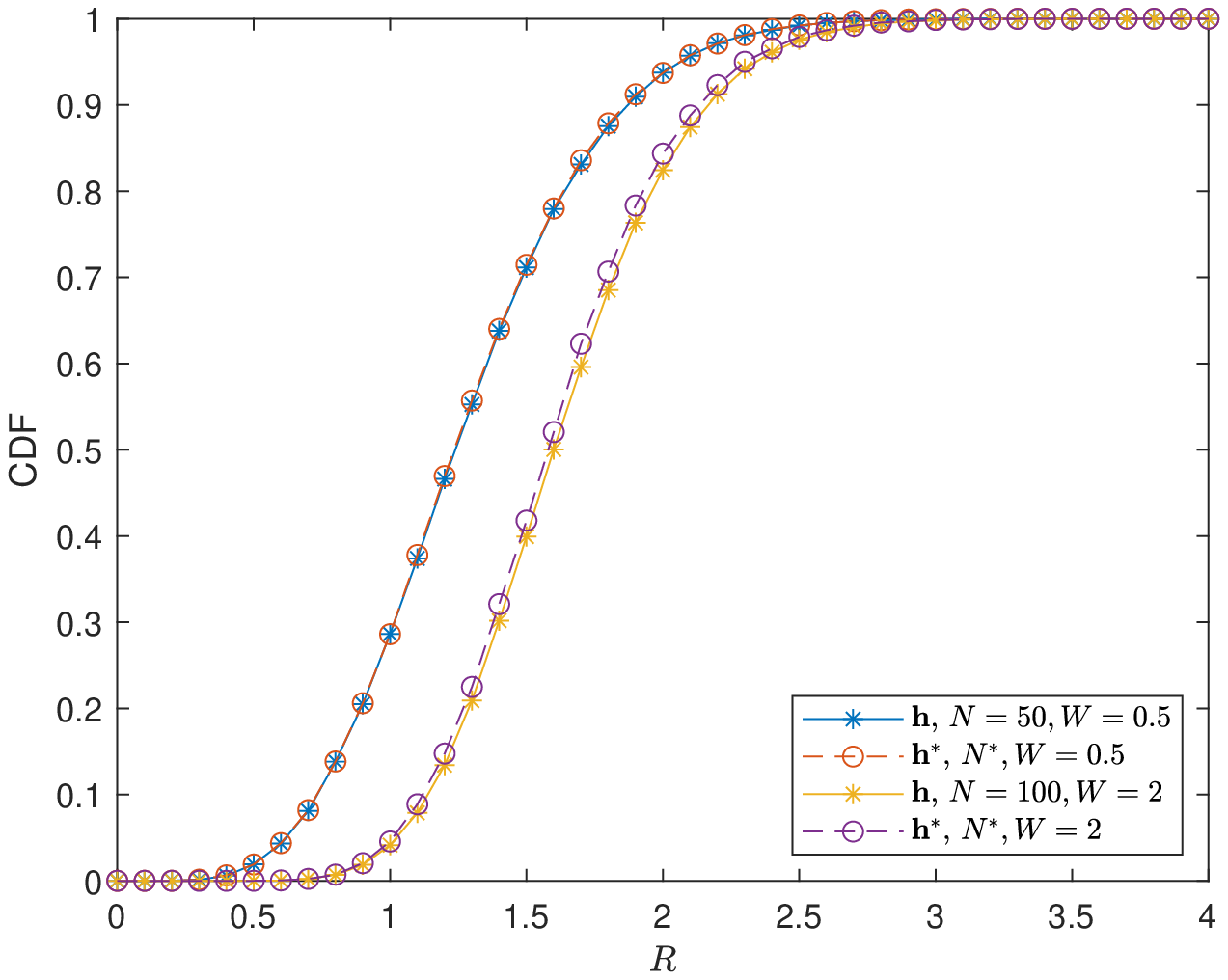} 
\par\end{centering}
}
\par\end{centering}
\caption{\label{fig:fig6}CDF between: a) $\boldsymbol{h}$ and $\tilde{\boldsymbol{h}}$;
b) $\boldsymbol{h}$ and $\boldsymbol{h}^{*}$.}
\end{figure}

Fig. \ref{fig:fig6}(a) presents the CDF of $\boldsymbol{h}$ and
$\tilde{\boldsymbol{h}}$ where we fix $R_{1}=\cdots=R_{N}=R$. In
the result, no significant variation is observed between $\boldsymbol{h}$
and $\tilde{\boldsymbol{h}}$ regardless of $R$, $N$ and $W$. This
is because the Frechet distance between the two distributions is always
near zero. This confirms Theorem \ref{thm:approx} and suggests that
one can always use $\tilde{\boldsymbol{h}}$ instead of $\boldsymbol{h}$.
In addition, Fig. \ref{fig:fig6}(b) shows the CDF of $\boldsymbol{h}$
and $\boldsymbol{h}^{*}$. Unlike the previous result, there is a
small gap between the two distributions as $W$ increases. Despite
having some gaps, the approximation is still fairly good. This result
verifies Theorem \ref{thm:singular}.
\begin{figure}
\begin{centering}
\includegraphics[scale=0.52]{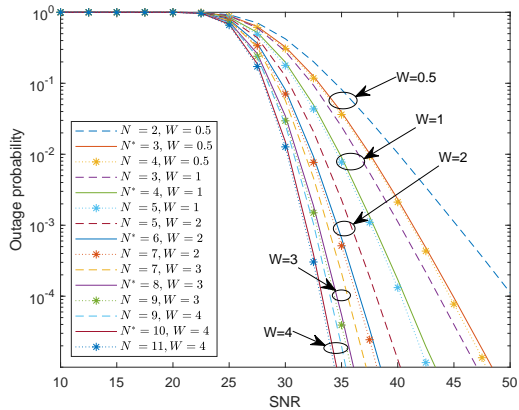} 
\par\end{centering}
\caption{\label{fig:fig7}Outage probability of suboptimal FAS.}
\end{figure}
\begin{table}
\caption{\label{table2}Parameter $N^{*}$ for different $W$ using algorithm
1 where $\varepsilon_{\text{tol}}=0.01$}

\centering{}%
\begin{tabular}{c||c|c|c|c|c}
\hline 
$\quad W\quad$ & 0.5 & 1 & 2 & 3 & 4\tabularnewline
\hline 
$\quad N^{*}\quad$ & 3 & 4 & 6 & 8 & 10\tabularnewline
\hline 
\end{tabular}
\end{table}

Next, we investigate the accuracy of Algorithm \ref{alg:algo1} and
the efficiency of the proposed suboptimal FAS. The parameter $N^{*}$
for different $W$ using Algorithm \ref{alg:algo1} is summarized
in Table \ref{table2}. As seen in Fig. \ref{fig:fig7}, the outage
probability of FAS with $N^{*}$ ports is promising. Specifically,
FAS with $N^{*}$ ports yields a significant improvement over FAS
with $N^{*}-1$ ports. Meanwhile, FAS with $N+1$ ports provides negligible
improvement over FAS with $N^{*}$ ports. Thus, we may use the suboptimal
FAS for an efficient performance.

\begin{figure}
\begin{centering}
\includegraphics[scale=0.52]{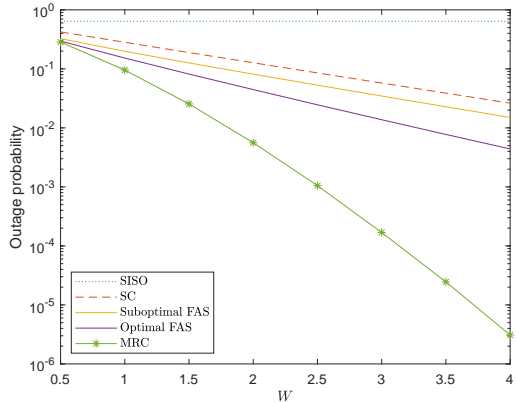} 
\par\end{centering}
\caption{\label{fig:fig8}Outage probability of suboptimal FAS vs. SISO, SC,
and MRC.}
\end{figure}

Finally in Fig. \ref{fig:fig8}, we compare the outage probability
of the proposed suboptimal FAS, the optimal FAS, the single antenna
(SISO) system, the $N$-branch SC system, and the $N$-branch MRC
system. In SC and MRC systems, we assume there are $N$ RF-chains
where each antenna has to be at least $\frac{\lambda}{2}$ apart and
their spatial correlations are considered. Note that
MRC has $N$ active RF-chains. Results show that the proposed suboptimal
FAS outperforms SISO and SC systems. This improvement is due to the
ability of FAS switching to the best port within a finite $W$. 

In addition, MRC has the lowest outage probability
and it outperforms optimal FAS. This superiority is due to the power
gain where a larger number of active RF-chains (i.e., $\left\lfloor \frac{W}{0.5}\right\rfloor +1$)
is utilized in MRC while FAS has only one active RF-chain. Although
MRC is more superior than the suboptimal FAS, the latter can achieve
a similar performance as compared to the earlier when $W=0.5$. Yet,
it is important to recall that MRC has one additional RF-chain as
compared to the suboptimal FAS in this case. Thus, it will be very
interesting to compare the performance of MIMO-FAS and MIMO with the
same number of active RF-chains.

\section{Conclusions}

In this paper, we considered FAS and approximated its outage probability
and diversity gain in closed-form expressions. New meaningful insights
were obtained from the analytical results, and simulation results
were given to better understand the factors that limit the performance
of FAS. Our results showed that the performance of FAS strongly depends
on the spatial correlation matrix $\boldsymbol{J}$. Specifically,
increasing the ports beyond $N'$ yields no diversity gain in a point-to-point
setting. Instead, increasing $N$ causes the correlation matrix $\boldsymbol{J}$
to be ill-conditioned. To address this, one can either increase $W$
for a fixed $N$ or decrease $N$ for a fixed $W$. In addition, we
proposed a suboptimal FAS with $N^{*}$ ports. By fixing an appropriate
$\varepsilon_{\text{tol}}$, the proposed scheme enabled us to obtain
a significant gain over FAS with $N^{*}-1$ while it nearly achieved
the same performance as FAS with $N^{*}+1$ ports. Thus, the approximation
of $N^{*}$ is useful since a larger number of ports yields diminishing
gains and additional costs. Furthermore, $N^{*}$ can be used to approximate
the channels of FAS with $N$ ports if the correlation matrix $\boldsymbol{J}$
is near-singular. Last but not least, the proposed suboptimal FAS
outperforms SISO and SC systems but falls behind MRC due to having
a single active RF-chain. Nevertheless, it was discovered that suboptimal
FAS and MRC achieve similar performance when $W=0.5$. Thus, it would
be interesting to study the performance of MIMO-FAS and MIMO in the
future.

\section*{Appendix A: Approximated PDF of $\left|\boldsymbol{h}\right|$}

The exact PDF of $\left|\boldsymbol{h}\right|$ is first derived in
\cite{https://doi.org/10.1002/dac.3510,wiegand2019series,WIEGAND2020107664}.
In this paper, we employ similar steps and further approximate the
PDF of $\left|\boldsymbol{h}\right|$ by introducing $\boldsymbol{G}$:
an $N\times N$ matrix, using an accurate binomial theorem, and truncating
the infinite series to a finite one for ease of computation. According
to \cite{tse2005fundamentals}, the PDF of a circularly symmetric
complex Gaussian random variables is known as 
\begin{equation}
f\left(\boldsymbol{h}\right)=\frac{1}{\pi^{N}\text{{det}\ensuremath{\left(\boldsymbol{J}\right)}}}\exp\left\{ -\boldsymbol{h}^{H}\boldsymbol{J}^{-1}\boldsymbol{h}\right\} ,\label{eq:A1}
\end{equation}
where \textbf{$\boldsymbol{J}^{-1}=\frac{\boldsymbol{K}^{T}}{\text{{det}\ensuremath{\left(\boldsymbol{J}\right)}}}$}
via Crammer rule. Using \cite[ (7-8) \& (7-9)]{edition2002probability},
the PDF of (\ref{eq:A1}) in terms of its amplitude and phase can
be obtained as 
\begin{equation}
f_{\left|\boldsymbol{h}\right|,\boldsymbol{\theta}}\left(\left|h_{1}\right|,\theta_{1},\ldots,\left|h_{N}\right|,\theta_{N}\right)=\eta\prod_{t=1}^{T}\exp\left\{ \zeta_{t}\cos\left(\bar{\theta}_{t}\right)\right\} ,\label{eq:A2}
\end{equation}
where $\eta=\frac{\stackrel[n=1]{N}{\prod}\left|h_{n}\right|}{\pi^{N}\text{{det}\ensuremath{\left(\boldsymbol{J}\right)}}}\exp\left\{ -\frac{\sum_{n=1}^{N}\left|h_{n}\right|^{2}K_{n,n}}{\text{{det}\ensuremath{\left(\boldsymbol{J}\right)}}}\right\} $,
$T=\frac{N\left(N-1\right)}{2}$, $\zeta_{t}=-\frac{2K_{m,n}\left|h_{n}\right|\left|h_{m}\right|}{\text{{det}\ensuremath{\left(\boldsymbol{J}\right)}}}$
and $\bar{\theta}_{t}=\theta_{n}-\theta_{m}$. Throughout this paper,
we use the mapping function $t=n+\left(m-1\right)N-\frac{m\left(m+1\right)}{2}$,
$m<n$, while $\left(m,n\right)$ can be obtained from $t$ by setting
$m=\min\:m'\in\mathbb{Z}$ subject to $\sum_{i=1}^{m'}\left(N-i\right)>t$
and $n=t-\left(m-1\right)N+\frac{m\left(m+1\right)}{2}$.

Integrating (\ref{eq:A2}) w.r.t. $\theta_{n},\forall n$ over $\left[0,2\pi\right]$,
we have 
\begin{align}
 & f_{\left|\boldsymbol{h}\right|}\left(\left|h_{1}\right|,\ldots,\left|h_{N}\right|\right)\nonumber \\
=~ & \int_{0}^{2\pi}\cdots\int_{0}^{2\pi}f\left(\left|h_{1}\right|,\theta_{1},\ldots,\left|h_{N}\right|,\theta_{N}\right)d\theta_{1}\ldots d\theta_{N}\label{eq:A3}\\
\overset{\left(a\right)}{=}~ & \eta\int_{0}^{2\pi}\cdots\int_{0}^{2\pi}\prod_{t=1}^{T}\sum_{s_{t}=0}^{\infty}\frac{\zeta_{t}^{s_{t}}}{s_{t}!}\cos\left(\bar{\theta}_{t}\right)^{s_{t}}d\theta_{1}\ldots d\theta_{N}\label{eq:A4}\\
\overset{\left(b\right)}{=}~ & \eta\sum_{s_{1}=0}^{\infty}\sum_{s_{2}=0}^{s_{1}}\ldots\sum_{s_{T}=0}^{s_{T-1}}\prod_{t=1}^{T}\beta\left(t,s_{t}^{*}\right)\int_{0}^{2\pi}\cdots\int_{0}^{2\pi}\cos\left(\bar{\theta}_{t}\right)^{s_{t}^{*}}d\theta_{1}\ldots d\theta_{N}\label{eq:A5}\\
\overset{\left(c\right)}{=}~ & \eta\sum_{s_{1}=0}^{\infty}\sum_{s_{2}=0}^{s_{1}}\ldots\sum_{s_{T}=0}^{s_{T-1}}\left(\frac{1}{2}\right)^{\sum_{t=1}^{T}s_{t}^{*}}\prod_{t=1}^{T}\beta\left(t,s_{t}^{*}\right)\times\label{eq:A6}\\
 & \int_{0}^{2\pi}\cdots\int_{0}^{2\pi}\prod_{t=1}^{T}\left(\exp\left\{ j\bar{\theta}_{t}\right\} +\exp\left\{ -j\bar{\theta}_{t}\right\} \right)^{s_{t}^{*}}d\theta_{1}\ldots d\theta_{N}\nonumber \\
\overset{\left(d\right)}{=}~ & \eta\sum_{s_{1}=0}^{\infty}\ldots\sum_{s_{T}=0}^{s_{T-1}}\left(\frac{1}{2}\right)^{\sum_{t=1}^{T}s_{t}^{*}}\prod_{t=1}^{T}\beta\left(t,s_{t}^{*}\right)\sum_{\boldsymbol{v}\in\mathcal{V}}\prod_{t=1}^{T}\left(\begin{array}{c}
s_{t}^{*}\\
v_{t}
\end{array}\right)\times\label{eq:A7}\\
 & \int_{0}^{2\pi}\cdots\int_{0}^{2\pi}\exp\left\{ j\sum_{t=1}^{T}\gamma_{t}\bar{\theta}_{t}\right\} d\theta_{1}\ldots d\theta_{N},\nonumber 
\end{align}
where (\ref{eq:A4}) is obtained by using $\exp\left\{ x\right\} =\sum_{s=0}^{\infty}\frac{x^{s}}{s!}$
and (\ref{eq:A5}) is obtained using Cauchy product of power series
where $\beta\left(t,s_{t}^{*}\right)\triangleq\frac{\zeta_{t}^{s_{t}^{*}}}{s_{t}^{*}!}$
and $s_{t}^{*}=s_{t}-s_{t+1}$ with $s_{T+1}=0$. Furthermore, (\ref{eq:A6})
is obtained using $\cos\left(x\right)=\frac{\exp\left(jx\right)+\exp\left(-jx\right)}{2}$
and (\ref{eq:A7}) is obtained using binomial theorem where $\boldsymbol{v}=\left[v_{1},\ldots,v_{T}\right]^{T}$,
$\mathcal{V}$ denotes the set of all the possible permutations and
$\gamma_{t}=2v_{t}-s_{t}^{*}\in\mathbb{Z}$.

Note that $\int_{0}^{2\pi}\cdots\int_{0}^{2\pi}\exp\left\{ j\sum_{t=1}^{T}\gamma_{t}\bar{\theta}_{t}\right\} d\theta_{1}\ldots d\theta_{N}=\left(2\pi\right)^{N}$
if and only if $\sum_{t=1}^{T}\gamma_{t}\bar{\theta}_{t}=0$, and
otherwise zero. Therefore, we introduce a new matrix $\boldsymbol{G}$
as defined in (\ref{eq:11}) and the matrix $\bar{\boldsymbol{\Theta}}$
given by 
\begin{align}
\bar{\boldsymbol{\Theta}} & =\left[\begin{array}{ccccc}
0 & \bar{\theta}_{1} & \bar{\theta}_{2} & \ldots & \bar{\theta}_{N-1}\\
 &  & \bar{\theta}_{N} & \ldots & \bar{\theta}_{2N-3}\\
\vdots &  & \ddots &  & \vdots\\
 &  &  &  & \bar{\theta}_{T}\\
0 &  & \ldots &  & 0
\end{array}\right]=\left[\begin{array}{ccccc}
0 & \theta_{2}-\theta_{1} & \theta_{3}-\theta_{1} & \ldots & \theta_{N}-\theta_{1}\\
 &  & \theta_{3}-\theta_{2} & \ldots & \theta_{N}-\theta_{2}\\
\vdots &  & \ddots &  & \vdots\\
 &  &  &  & \theta_{N}-\theta_{N-1}\\
0 &  & \ldots &  & 0
\end{array}\right].\label{eq:A8}
\end{align}
Using $\bar{\boldsymbol{\Theta}}$ and $\boldsymbol{G}$, we can easily
integrate (\ref{eq:A7}) w.r.t. to $\theta_{i}$ by taking the sum
of the same entries of $\boldsymbol{G}$ as that of $\bar{\boldsymbol{\Theta}}$
with $\theta_{i}$, i.e., $\Delta_{i}=\sum_{n=1}^{N}G_{i,n}-\sum_{n=1}^{N}G_{n,i}-G_{i,i}$.
Therefore, (\ref{eq:A7}) leads to 
\begin{align}
\left(\ref{eq:A7}\right)=~ & \eta\sum_{s_{1}=0}^{\infty}\sum_{s_{2}=0}^{s_{1}}\ldots\sum_{s_{T}=0}^{s_{T-1}}\left(\frac{1}{2}\right)^{\sum_{t=1}^{T}s_{t}^{*}}\prod_{t=1}^{T}\beta\left(t,s_{t}^{*}\right)\sum_{\boldsymbol{v}\in\mathcal{V}}\left[\prod_{t=1}^{T}\left(\begin{array}{c}
s_{t}^{*}\\
v_{t}
\end{array}\right)\right]\left[\left(2\pi\right)^{N}\prod_{i=1}^{N}\boldsymbol{1}_{\left\{ \Delta_{i}=0\right\} }\right]\label{eq:A9}\\
\overset{\left(a\right)}{\approx}~ & \eta\stackrel[s_{1}=0]{s_{0}}{\sum}\sum_{s_{2}=0}^{s_{1}}\ldots\sum_{s_{T}=0}^{s_{T-1}}\left(\frac{1}{2}\right)^{\sum_{t=1}^{T}s_{t}^{*}}\prod_{t=1}^{T}\beta\left(t,s_{t}^{*}\right)\sum_{\boldsymbol{v}\in\mathcal{V}}\left[\prod_{t=1}^{T}\left(\begin{array}{c}
s_{t}^{*}\\
v_{t}
\end{array}\right)\right]\left[\left(2\pi\right)^{N}\prod_{i=1}^{N}\boldsymbol{1}_{\left\{ \Delta_{i}=0\right\} }\right],\label{eq:A10}
\end{align}
where $(a)$ can be obtained using the facts that
$\left(\frac{1}{2}\right)^{\sum_{t=1}^{T}s_{t}^{*}}$ is monotonically
decreasing in each summation term and $\beta\left(t,s_{t}^{*}\right)\approx0$
if $s_{t}^{*}$ is sufficiently large.

\section*{Appendix B: Approximated CDF of $\left|\boldsymbol{h}\right|$}

Using (\ref{eq:10}), the CDF of $\left|\boldsymbol{h}\right|$ can
be obtained as 
\begin{align}
F\left(R_{1},\ldots,R_{N}\right)\approx & \int_{0}^{R_{1}}\ldots\int_{0}^{R_{N}}f_{\left|\boldsymbol{h}\right|}\left(\left|h_{1}\right|,\ldots,\left|h_{N}\right|\right)d\left|h_{1}\right|\cdots d\left|h_{N}\right|\label{eq:B1}\\
= & \stackrel[s_{1}=0]{s_{0}}{\sum}\sum_{s_{2}=0}^{s_{1}}\ldots\sum_{s_{T}=0}^{s_{T-1}}\frac{g\left(\boldsymbol{s}^{*}\right)}{\pi^{N}\text{{det}\ensuremath{\left(\boldsymbol{J}\right)}}}\prod_{t=1}^{T}\frac{\left(-2K_{m,n}\right)^{s_{t}^{*}}}{s_{t}^{*}!\text{{det}\ensuremath{\left(\boldsymbol{J}\right)}}^{s_{t}^{*}}}\int_{0}^{R_{1}}\ldots\int_{0}^{R_{N}}\times\label{eq:B2}\\
 & \stackrel[n=1]{N}{\prod}\left|h_{n}\right|\prod_{n=1}^{N}\prod_{m<n}^{N}\left|h_{n}\right|^{s_{n}^{*}}\left|h_{m}\right|^{s_{m}^{*}}\exp\left\{ -\frac{\sum_{n=1}^{N}\left|h_{n}\right|^{2}K_{n,n}}{\text{{det}\ensuremath{\left(\boldsymbol{J}\right)}}}\right\} d\left|h_{1}\right|\cdots d\left|h_{N}\right|\nonumber \\
= & \stackrel[s_{1}=0]{s_{0}}{\sum}\sum_{s_{2}=0}^{s_{1}}\ldots\sum_{s_{T}=0}^{s_{T-1}}\frac{g\left(\boldsymbol{s}^{*}\right)}{\pi^{N}\text{{det}\ensuremath{\left(\boldsymbol{J}\right)}}}\prod_{t=1}^{T}\frac{\left(-2K_{m,n}\right)^{s_{t}^{*}}}{s_{t}^{*}!\text{{det}\ensuremath{\left(\boldsymbol{J}\right)}}^{s_{t}^{*}}}\times\label{eq:B3}\\
 & \stackrel[n=1]{N}{\prod}\int_{0}^{R_{n}}\left|h_{n}\right|^{\bar{s}_{n}+1}\exp\left\{ -\frac{\left|h_{n}\right|^{2}K_{n,n}}{\text{{det}\ensuremath{\left(\boldsymbol{J}\right)}}}\right\} d\left|h_{n}\right|\nonumber \\
= & \stackrel[s_{1}=0]{s_{0}}{\sum}\sum_{s_{2}=0}^{s_{1}}\ldots\sum_{s_{T}=0}^{s_{T-1}}\frac{g\left(\boldsymbol{s}^{*}\right)}{\pi^{N}\text{{det}\ensuremath{\left(\boldsymbol{J}\right)}}}\prod_{t=1}^{T}\frac{\left(-K_{m,n}\right)^{s_{t}^{*}}}{s_{t}^{*}!\text{{det}\ensuremath{\left(\boldsymbol{J}\right)}}^{s_{t}^{*}}}\times\label{eq:B4}\\
 & \stackrel[n=1]{N}{\prod}\frac{1}{2}\left(\frac{\text{\ensuremath{K_{n,n}}}}{\text{{det}\ensuremath{\left(\boldsymbol{J}\right)}}}\right)^{-\frac{\bar{s}_{n}}{2}-\frac{1}{2}}\left[\Gamma\left(\frac{1+\bar{s}_{n}}{2}\right)-\Gamma\left(\frac{1+\bar{s}_{n}}{2},\frac{\text{\ensuremath{K_{n,n}}}R_{n}^{2}}{\text{{det}\ensuremath{\left(\boldsymbol{J}\right)}}}\right)\right],\nonumber 
\end{align}
where 
\begin{equation}
g\left(\boldsymbol{s}^{*}\right)=\left(\frac{1}{2}\right)^{\sum_{t=1}^{T}s_{t}^{*}}\sum_{\boldsymbol{v}\in\mathcal{V}}\left[\prod_{t=1}^{T}\left(\begin{array}{c}
s_{t}^{*}\\
v_{t}
\end{array}\right)\right]\left(2\pi\right)^{N}\prod_{i=1}^{N}\boldsymbol{1}_{\left\{ \Delta_{i}=0\right\} },\label{eq:B5}
\end{equation}
and $\bar{s}_{n}$ is the sum of $s_{t}^{*}$ affecting $\left(\left|h_{n}\right|\left|h_{m}\right|\right)^{s_{t}^{*}}$.
To compute $\bar{s}_{n}$ , let us introduce a new matrix 
\begin{equation}
\boldsymbol{S}^{*}=\left[\begin{array}{ccccc}
0 & s_{1}^{*} & s_{2}^{*} & \ldots & s_{N-1}^{*}\\
 &  & s_{N}^{*} & \ldots & s_{2N-3}^{*}\\
\vdots &  & \ddots &  & \vdots\\
 &  &  &  & s_{T}^{*}\\
0 &  & \ldots &  & 0
\end{array}\right].\label{eq:B6}
\end{equation}
Using (\ref{eq:B6}), we have $\bar{s}_{n}=\sum_{i=1}^{N}S_{n,i}^{*}+\sum_{i=1}^{n-1}S_{i,n}^{*}+1$
such that $S_{i,n}^{*}$ is the $\left(i,n\right)$-th entry of $\boldsymbol{S}^{*}$.

\section*{Appendix C: Outage Probability at High SNR}

According to \cite{1221802}, the outage probability of a wireless
communication system at high SNR can be obtained via the PDF of its
fading channels. In particular, suppose the PDF of the channels at
high SNR can be approximated as 
\begin{equation}
f_{\left|h_{\text{FAS}}\right|}\left(\Omega\right)=2\xi\Omega^{2M+1}+o\left(\Omega^{2M+1}\right).\label{eq:C1}
\end{equation}
Then the outage probability at high SNR is found as 
\begin{equation}
\mathcal{\mathbb{P}}\left\{ \left|h_{\text{{FAS}}}\right|<\varOmega\right\} =\frac{\xi}{M+1}\varOmega^{2\left(M+1\right)}+o\left(\frac{1}{SNR^{M+1}}\right).\label{eq:C2}
\end{equation}

Before approximating the PDF of FAS at high SNR, we highlight that
the PDF of (\ref{eq:A1}) in terms of its amplitude and phase can
be rewritten as 
\begin{equation}
f_{\left|\boldsymbol{h}\right|,\boldsymbol{\theta}}\left(\left|h_{1}\right|,\theta_{1},\ldots,\left|h_{N}\right|,\theta_{N}\right)=\stackrel[n=1]{N}{\prod}\frac{\left|h_{n}\right|H_{n}}{\pi^{N}\text{{det}\ensuremath{\left(\boldsymbol{J}\right)}}},\label{eq:C3}
\end{equation}
where 
\begin{align}
H_{n} & =\exp\Bigg\{-\frac{K_{n,n}\left|h_{n}\right|^{2}}{\text{{det}\ensuremath{\left(\boldsymbol{J}\right)}}}-\frac{2\sum_{m=n+1}^{N}K_{m,n}\left|h_{n}\right|\left|h_{m}\right|\cos\left(\theta_{n}-\theta_{m}\right)}{\text{{det}\ensuremath{\left(\boldsymbol{J}\right)}}}\Bigg\}.\label{eq:C4}
\end{align}
Using (\ref{eq:C3}), the approximated PDF of FAS at high SNR can
be derived as 
\begin{align}
f_{\left|h_{\text{FAS}}\right|}\left(\Omega\right)= & \frac{\partial F_{\left|h_{\text{FAS}}\right|}\left(\Omega\right)}{\partial\Omega}\label{eq:C5}\\
\overset{\left(a\right)}{=} & N\int_{0}^{\Omega}\ldots\int_{0}^{\Omega}\int_{0}^{2\pi}\cdots\int_{0}^{2\pi}f_{\left|\boldsymbol{h}\right|,\boldsymbol{\theta}}\left(\left|h_{1}\right|,\theta_{1},\ldots,\left|h_{N-1}\right|,\theta_{N-1},\Omega,\theta_{N}\right)\times\label{eq:C6}\\
 & d\left|h_{1}\right|\cdots d\left|h_{N-1}\right|d\theta_{1}\ldots d\theta_{N}\nonumber \\
\overset{\left(b\right)}{=} & \frac{N\Omega}{\pi^{N}\text{{det}\ensuremath{\left(\boldsymbol{J}\right)}}}\int_{0}^{2\pi}\cdots\int_{0}^{2\pi}H_{N}\int_{0}^{\varOmega}\left|h_{N-1}\right|\bigg(H_{N-1}\times\ldots\label{eq:C7}\\
 & \left(\int_{0}^{\varOmega}\left|h_{2}\right|H_{2}\left(\int_{0}^{\varOmega}\left|h_{1}\right|H_{1}d\left|h_{1}\right|\right)d\left|h_{2}\right|\right)d\left|h_{N-1}\right|\Bigg)d\theta_{1}\ldots d\theta_{N},\nonumber 
\end{align}
where $(a)$ is obtained using Leibniz integral and $(b)$ is obtained
using (\ref{eq:C3}).

According to \cite{5464234}, the term $\int_{0}^{\varOmega}\left|h_{n}\right|H_{n}d\left|h_{n}\right|$
can be solved by applying Taylor series approximation at around zero.
Specifically, we have 
\begin{align}
\int_{0}^{\varOmega}\left|h_{n}\right|H_{n}d\left|h_{n}\right|=\frac{\varOmega^{2}}{2}+o\left(\varOmega^{2}\right),n=\left\{ 1,\ldots,N-1\right\} \label{eq:C8}
\end{align}
and the Taylor series approximation of $H_{N}$ at zero is 
\begin{align}
H_{N} & =1+o\left(1\right).\label{eq:C9}
\end{align}
Substituting (\ref{eq:C8}) and (\ref{eq:C9}) into (\ref{eq:C7}),
we have 
\begin{align}
 & f_{\left|h_{\text{FAS}}\right|}\left(\Omega\right)\nonumber \\
= & \frac{N\Omega}{\pi^{N}\text{{det}\ensuremath{\left(\boldsymbol{J}\right)}}}\left[\frac{\varOmega^{2}}{2}+o\left(\varOmega^{2}\right)\right]^{N-1}\int_{0}^{2\pi}\cdots\int_{0}^{2\pi}d\theta_{1}\ldots d\theta_{N}\label{eq:C10}\\
= & \frac{2N}{\text{{det}\ensuremath{\left(\boldsymbol{J}\right)}}}\varOmega^{2N-1}+o\left(\varOmega^{2N-1}\right).\label{eq:C11}
\end{align}
Comparing (\ref{eq:C11}) to (\ref{eq:C1}), we have $M=N-1$ and
$\xi=\frac{N}{\text{{det}\ensuremath{\left(\boldsymbol{J}\right)}}}$.
Applying (\ref{eq:C2}), we have 
\begin{equation}
\mathcal{\mathbb{P}}\left\{ \left|h_{\text{{FAS}}}\right|<\varOmega\right\} \approx\frac{1}{\text{{det}\ensuremath{\left(\boldsymbol{J}\right)}}}\varOmega^{2N}+o\left(\frac{1}{SNR^{N}}\right).\label{eq:C12}
\end{equation}

\section*{Appendix D: Diversity Gain of FAS}

Let us consider the case where $W\rightarrow\infty$. According to
\cite{1221802}, the diversity gain of a wireless communication system
can be obtained via the PDF of its fading channels at high SNR. Specifically,
suppose the PDF of the channels at high SNR can be approximated as
in (\ref{eq:C1}). Then diversity gain of such system is given by
\begin{equation}
D=M+1.\label{eq:D1}
\end{equation}
In Appendix C, we have $M=N-1$. Thus, it is straightforward that
the diversity gain of FAS as $W\rightarrow\infty$ is $N$. Nevertheless,
if $W$ is finite, $\boldsymbol{J}$ might be near to being singular.
To see this, let us consider FAS with $N\rightarrow\infty$ ports
within a finite $W$ where each port is equally separated, and they
are indexed as $1,2,\ldots$. Without loss of generality, let us focus
on two ports: the $n$-th and $(n+1)$-th port. The correlation between
the $n$-th port and $(n+1)$-th port is $\boldsymbol{J}_{n,n+1}=\underset{N\rightarrow\infty}{\lim}\sigma^{2}J_{0}\left(2\pi\frac{1}{N-1}W\right)=\sigma^{2}J_{0}\left(0\right)$,
and we have $h_{n+1}=h_{n}$. Thus, the joint CDF of $h_{n}$ and
$h_{n+1}$ is $F_{h_{n},h_{n+1}}\left(g_{1},g_{2}\right)=F_{h_{n}}\left(\min\left\{ g_{1},g_{2}\right\} \right)$,
which implies that they reduce to singularity. Since there are many
such ports, we can use a finite $N'$ ports to approximate the channels
of FAS with $N$ ports, where $N'$ is the numerical rank of $\boldsymbol{J}'$
such that $\boldsymbol{J}'$ is covariance matrix as defined in (\ref{eq:2})
with $N\rightarrow\infty$ for a fixed $W$. As a result, the diversity
gain of FAS is approximately limited by $\min\left\{ N,N'\right\} $.
If $N$ is large, the same observation can be obtained. To remove
the nearly-dependent entries of $\boldsymbol{J}$, one may employ
rank-revealing QR factorization \cite{golub1965numerical} or Gauss-Jordan
elimination with a given tolerance.

\bibliographystyle{IEEEtran}
\bibliography{Fluid_Antenna}

\end{document}